\documentclass[useAMS,usenatbib,letterpaper]{mn2e}
\usepackage[pass]{geometry}
\usepackage{natbib}
\usepackage{amssymb,amsmath}
\usepackage[ruled,linesnumbered]{algorithm2e}
\usepackage{bm}
\usepackage{relsize}
\usepackage{graphicx}
\usepackage{enumitem} 
\usepackage{verbatim}
\usepackage{color,hyperref}
\usepackage{lastpage}
\definecolor{linkcolor}{rgb}{0,0,0.25}
\hypersetup{
  colorlinks=true,        
  linkcolor=linkcolor,    
  citecolor=linkcolor,    
  filecolor=linkcolor,    
  urlcolor=linkcolor      
}
\newcommand{\etal}{et al.}
\newcommand{\dd}{\mathrm{d}}
\newcommand{\eg}{e.g.}

\newcommand{\Eqnname}{Equation}
\newcommand{\equationname}{\Eqnname}

\newcommand{\figurename}{Figure}

\newcommand{\sectionname}{$\mathsection$}
\newcommand{\appendixname}{Appendix}

\renewcommand{\vec}[1]{\ensuremath{\mathbf{#1}}}

\newcommand{\Myr}{\ensuremath{\,\mathrm{Myr}}}
\newcommand{\Gyr}{\ensuremath{\,\mathrm{Gyr}}}
\newcommand{\kpc}{\ensuremath{\,\mathrm{kpc}}}
\newcommand{\pc}{\ensuremath{\,\mathrm{pc}}}
\newcommand{\kms}{\ensuremath{\,\mathrm{km\ s}^{-1}}}

\newcommand{\inv}{\ensuremath{^{-1}}}

\newcommand{\dens}{\ensuremath{\nu}}
\newcommand{\zsun}{\ensuremath{z_\odot}}
\newcommand{\vz}{\ensuremath{v_z}}
\newcommand{\vzi}{\ensuremath{v_{z,i}}}
\newcommand{\zi}{\ensuremath{z_i}}
\newcommand{\wi}{\ensuremath{w_i}}
\newcommand{\yi}{\ensuremath{y_i}}

\newcommand{\Ei}{\ensuremath{E_i}}
\newcommand{\Ji}{\ensuremath{J_i}}
\newcommand{\Ai}{\ensuremath{A_i}}
\newcommand{\phii}{\ensuremath{\phi_i}}
\newcommand{\zobs}{\ensuremath{\tilde{z}}}
\newcommand{\eps}{\ensuremath{\epsilon}}

\def\aj{AJ}
\def\apj{ApJ}

\def\mnras{MNRAS}
\def\nat{Nature}
\def\aap{A \& A}

\title[Made-to-measure modeling of observed galaxy dynamics]{Made-to-measure modeling of observed galaxy dynamics}
\author[Bovy, Kawata, \& Hunt]{Jo Bovy$^{1,2}$\thanks{E-mail: bovy@astro.utoronto.ca}\thanks{Alfred~P.~Sloan~Fellow}, Daisuke Kawata$^3$, \& Jason A.~S.~Hunt$^4$\\
$^1$Department of Astronomy and Astrophysics, University of Toronto, 50 St. George Street, Toronto, ON M5S 3H4, Canada\\
$^2$Center for Computational Astrophysics, Flatiron Institute, 162 5th Ave, New York, NY 10010, USA\\
$^3$Mullard Space Science Laboratory, University College London, Holmbury St. Mary, Dorking, Surrey, RH5 6NT, United Kingdom\\
$^4$Dunlap Institute for Astronomy and Astrophysics, University of Toronto, 50 St. George Street, Toronto, Ontario, M5S 3H4, Canada}
\pagerange{\pageref{firstpage}--\pageref{LastPage}} \pubyear{2017}
\date{6 September 2017}

\begin{document}
\maketitle
\label{firstpage}
\begin{abstract}
  Among dynamical modeling techniques, the made-to-measure (M2M)
  method for modeling steady-state systems is among the most flexible,
  allowing non-parametric distribution functions in complex
  gravitational potentials to be modeled efficiently using $N$-body
  particles. Here we propose and test various improvements to the
  standard M2M method for modeling observed data, illustrated using
  the simple setup of a one-dimensional harmonic oscillator. We
  demonstrate that nuisance parameters describing the modeled system's
  orientation with respect to the observer---\eg, an external galaxy's
  inclination or the Sun's position in the Milky Way--- as well as the
  parameters of an external gravitational field can be optimized
  simultaneously with the particle weights. We develop a method for
  sampling from the high-dimensional uncertainty distribution of the
  particle weights. We combine this in a Gibbs sampler with samplers
  for the nuisance and potential parameters to explore the uncertainty
  distribution of the full set of parameters. We illustrate our M2M
  improvements by modeling the vertical density and kinematics of
  F-type stars in \emph{Gaia} DR1. The novel M2M method proposed here
  allows full probabilistic modeling of steady-state dynamical
  systems, allowing uncertainties on the non-parametric distribution
  function and on nuisance parameters to be taken into account when
  constraining the dark and baryonic masses of stellar systems.
\end{abstract}

\section{Introduction}

Constraining the orbital structure and mass distribution of
astrophysical systems through dynamical modeling is one of the
fundamental ways to learn about the dark-matter and baryonic
distribution in external galaxies
\citep[\eg,][]{Rix97a,Cappellari12a}, supermassive black holes at the
centers of galaxies \citep[\eg,][]{Magorrian98a}, and the mass
distribution of the Milky Way \citep[\eg,][]{Bovy13a}, to name but a
few. Of particular interest are systems---such as galaxies or star
clusters---that may be assumed to be in a steady state. Many
techniques have been proposed to model such systems, typically
combining the steady-state assumption with further assumptions about
the orbital structure (\eg, the velocity anisotropy) or symmetry (\eg,
spherical or axisymmetric) of the system. The simplest among these
techniques are those based on moments of the collisionless Boltzmann
equation, \eg, the Jeans equations, which despite their restrictive
assumptions remain a useful tool for interpreting data
\citep[\eg,][]{Cappellari13a}. A second class of techniques directly
uses parameterized distribution functions (DFs) that satisfy the
collisionless Boltzmann equation by only depending on integrals of the
motion. While restricted to gravitational potentials for which such
integrals can be computed, this class of models has reached a high
level of sophistication, especially in the Milky Way
\citep[\eg,][]{Binney10a,Bovy13a,Trick16a}. A third class of methods
eschews parameterized DFs, but rather builds a steady-state model in a
fixed gravitational potential from a large number of orbit building
blocks with weights determined by fitting a set of constraints
\citep{Schwarzschild79a,Schwarzschild93a}.

\citet{Syer96a} proposed a method known as made-to-measure (M2M)
modeling that is closely related to orbit-based modeling. In M2M, the
DF is represented not by entire orbits but instead by a set of
$N$-body particles with positions and velocities
$(\vec{x}_i,\vec{v}_i)$ and weights $w_i$. They demonstrated that a
steady-state solution to a set of constraints on the phase-space
distribution (expressed as a $\chi^2$,the mean-squared difference
between the model and the data) can be obtained by slowly adjusting
the weights of each particle in the direction of decreasing $\chi^2$
while integrating the orbits of the particles. The advantages of this
particle-based technique over orbit-based methods are that only the
current snapshot needs to be stored in memory rather than entire
orbits, that the $N$-body particles can contribute a self-consistent
part of the gravitational potential, and that one ends up with an
actual sampling from the steady-state DF. The latter makes M2M also an
ideal technique for initializing $N$-body simulations
\citep[\eg,][]{Dehnen09a,Malvido15a}.

Since its original description, various improvements have been made to
the basic M2M setup, such as allowing for observational uncertainties
and kinematic data in the constraints \citep{DeLorenzi07a,Long10a},
integrating particles on individual time scales for problems with a
range of orbital frequencies \citep{Dehnen09a}, improvements in the
smoothing applied to the model \citep{Dehnen09a}, and allowing data
for individual stars as constraints
\citep{Hunt13a,Hunt13b,Hunt14a}. As currently conceived, M2M modeling
applies to the particle weights only and any other parameter
describing the system is held fixed during the optimization. This
includes nuisance parameters describing the modeled system's
orientation with respect to the observer, for example, the inclination
of an external galaxy or the Sun's distance to the Galactic center for
Milky-Way applications, and the parameters of the external
gravitational field. Furthermore, as methods for modeling observed
data both Schwarzschild and M2M modeling remain problematic in that
they are fundamentally optimization algorithms that do not take into
account the uncertainties in the DF resulting from the strong
degeneracies among the large number of orbit or particle weights
\citep{Magorrian06a}. For obtaining the best constraints from a given
set of observables, a fully probabilistic treatment is warranted that
samples from the full uncertainty distribution for the particle
weights, nuisance parameters, and the parameters describing the
potential. In this paper we extend the basic M2M modeling framework to
optimize for nuisance and potential parameters simultaneously with the
particle weights and we introduce sampling methods to sample the
uncertainty distribution of all parameters.

The outline of this paper is as follows. In
\sectionname~\ref{sec:mock} we describe the simple, one-dimensional
setup that we use as a toy problem: modeling an isothermal population
in a external harmonic-oscillator potential. We describe the standard
M2M method in \sectionname~\ref{sec:standard}. In
\sectionname~\ref{sec:wunc} we discuss how to sample from the
uncertainty distribution of the particle weights. We show how one can
optimize the value of the nuisance parameters at the same time as the
values of the particle weights in \sectionname~\ref{sec:nuisance} and
give a Markov Chain Monte Carlo (MCMC) algorithm to sample both the
particle weights and the nuisance parameters. In
\sectionname~\ref{sec:potential}, we discuss how we can also optimize
the value of the parameters describing the external gravitational
potential simultaneously with the particle weights and the nuisance
parameters and present an MCMC algorithm for sampling all
parameters. To illustrate how the M2M improvements perform for real data,
we apply the new M2M algorithm to data on the density and kinematics
of F stars in \emph{Gaia} DR1 in \sectionname~\ref{sec:gaia}. We
discuss various aspects of this novel M2M method and avenues for
future work in \sectionname~\ref{sec:discussion} and present our
conclusions in \sectionname~\ref{sec:conclusion}.

\section{Harmonic-oscillator M2M: a simple testbed for M2M modeling}\label{sec:mock}

To illustrate and test our modeling extensions of the basic M2M
algorithm below, we consider a one-dimensional system with the
gravitational potential of a harmonic oscillator (HO). This setup is
chosen for its simplicity; everything that we describe below applies
more generally to full, three-dimensional M2M modeling. This setup is
an ideal testbed for M2M modeling because (a) orbit integration is
analytic, (b) the DF corresponding to a given potential and a given
density is unique (thus, there is a well-defined unique solution to
the M2M problem; \eg, \citealt{Kuijken89a}), (c) it is easy to write
down simple DF models, and (d) running the M2M modeling in practice is
very fast. While simple, this model is a also semi-realistic,
approximate representation of the vertical dynamics in the solar
neighborhood close to the mid-plane and thus has some practical
applicability (see \sectionname~\ref{sec:gaia}). We ignore the
self-gravity of the M2M $N$-body particles and the potential is thus
assumed to be external and fixed. In this section, we describe the
basic notation, equations, and concepts of this model.

We denote the phase-space coordinates as $(z,\vz)$. The HO potential
is
\begin{equation}
  \Phi(z;\omega) = \frac{\omega^2\,z^2}{2}\,,
\end{equation}
specified by a single parameter $\omega$, the oscillator's
frequency. Orbit integration in the HO potential is analytic: orbits
are given by
\begin{align}\label{eq:zit}
  \zi(t) & = \phantom{-}\Ai\phantom{\,\omega}\,\cos\left(\omega\,t+\phii\right)\,,\\
  \vzi(t) & = -\Ai\,\omega\,\sin\left(\omega\,t+\phii\right)\,,\label{eq:vzit}
\end{align}
where
\begin{align}\label{eq:Ai}
  A_i  & = z_{\mathrm{max}} = \frac{\sqrt{2\,E_i}}{\omega} = \sqrt{z_i^2(0)+\frac{\vzi^2(0)}{\omega^2}}\,,\\
  \phii & = \arctan\!2(-\vzi(0)/\omega,\zi(0))\,,\label{eq:phii}
\end{align}
in which $(\zi(0),\vzi(0))$ is the initial phase-space position of an
orbit indexed by $i$ and $\arctan\!2$ is the arc-tangent function that
chooses the quadrant correctly. 

In this HO potential, we attempt to match a population drawn from a DF
given by
\begin{equation}
  f(z,\vz) \propto e^{-E/\sigma^2}\,,
\end{equation}
where $E = \omega^2\,z^2 / 2 + \vz^2/2$ is the energy and $\sigma$ is
the velocity dispersion parameter. This DF is isothermal---it has the
same velocity dispersion at all heights $\langle v^2\rangle =
\sigma^2$---and in a steady-state, because it is only a function of
the conserved energy $E$. The density distribution for this
distribution is
\begin{equation}
  \dens(z) \propto \exp\left(-\frac{\omega^2\,z^2}{2\,\sigma^2}\right)\,,
\end{equation}
which is a Gaussian distribution with a standard deviation $\sigma_\nu
= \sigma / \omega$. The velocity distribution at each $z$ is a
Gaussian with dispersion $\sigma$. Sampling orbits at initial
phase-space locations $(\zi(0),\vzi(0))$ from $f(z,\vz) \propto
e^{-E/\sigma^2}$ is simple: (i) sample $\Ei$ from the exponential
distribution and convert it to $\Ai$; (ii) sample $\phii$ uniformly
between $0$ and $2\pi$; (iii) convert $(\Ai,\phii)$ to $(\zi,\vzi)$.

To fit this DF using M2M below, we start with $(\zi,\vzi)$ drawn with
uniform weights \wi\ from an isothermal DF, but with a different
$\sigma$ from the true velocity dispersion: $f(z,\vz) \propto
e^{-E/\sigma_{\mathrm{in}}^2}$, with $\sigma_{\mathrm{in}}$ typically
$0.2$. It is then easy to see that the correct output particle weights
for a true velocity-dispersion parameter $\sigma_{\mathrm{target}}$
should be
\begin{equation}
  \wi \propto \exp\left( -\Ei\left[\frac{1}{\sigma_{\mathrm{target}}^2}-\frac{1}{\sigma_{\mathrm{in}}^2}\right]\right)\,,
\end{equation}
if the potential remains fixed. If the potential is adiabatically
changed from a HO potential with frequency $\omega_{\mathrm{in}}$ to
one with frequency $\omega_{\mathrm{target}}$ the correct output
particle weights are
\begin{equation}
  \wi \propto \exp\left( -\Ji\left[\frac{\omega_{\mathrm{target}}}{\sigma_{\mathrm{target}}^2}-\frac{\omega_{\mathrm{in}}}{\sigma_{\mathrm{in}}^2}\right]\right)\,,
\end{equation}
where $\Ji = \Ei/\omega$ is the action.

\section{Standard M2M modeling}\label{sec:standard}

We first describe the standard M2M case. Standard M2M models a
steady-state DF as a set of $N$ particles ($\zi,\vzi)$ indexed by $i$
orbiting in a fixed potential. Each particle has a weight $\wi$ that
is adjusted on-the-fly during orbit integration to fit a set of
constraints, like the density in bins, or the velocity dispersion. By
only adjusting the particle weights $\wi$ on timescales $\gg$ the
orbital timescale, an approximate equilibrium distribution is
obtained.

In practice, M2M maximizes an objective function $F$ that represents a
balance between reproducing the constraints, expressed as $\chi^2$
differences between data and model, and a penalty $S$ that disfavors
non-smooth DFs
\begin{equation}\label{eq:objective}
  F = S - \frac{1}{2}\sum_j \chi^2_j\,.
\end{equation}

Traditionally, the penalty $S$ is implemented through a
maximum-entropy constraint by setting
\begin{equation}\label{eq:entropy}
S = - \mu \sum_i w_i\left[\ln\left(w_i/\hat{w}_i\right)-1\right]\,,
\end{equation}
where $\hat{w}_i$ is a default set of particle weights. In the absence
of constraints, the entropy penalty prefers $\wi = \hat{w}_i$. The
parameter $\mu$ quantifies the strength of the penalty.

Constraints are expressed as a kernel applied to the DF $f(z,\vz)$:
\begin{equation}
  Y_j = \int \dd z\dd \vz \,K_j(z,\vz) f(z,\vz)\,
\end{equation}
which for the $N$-body snapshot is computed as
\begin{equation}
  y_j = \sum_i \wi K_j(\zi,\vzi)\,.
\end{equation}

To illustrate the standard M2M case, we use the density and the
density-weighted mean-squared velocity, both observed at a few points
indexed by $j$. The model density is given by
\begin{equation}\label{eq:modeldens}
  \dens(\zobs_j) = \sum_i \wi K^0(|\zobs_j+\zsun - \zi|;h)\,,
\end{equation}
where $K^0(r;h)$ is a kernel function with a width parameter $h$
that integrates to one ($\int \dd r\,K^0(r;h) = 1$) and we assume
that the observations are done as a function of $\zobs$, which is
measured with respect to the observer's position, located at
\zsun\ from the $z=0$ midplane (we give the specific kernel and
$\zobs$ used in this paper in \sectionname~\ref{sec:example}). In what
follows, we will abbreviate $K^0_j(\zi;h) \equiv
K^0(|\zobs_j+\zsun - \zi|;h)$. We assume that the density is
observed with a Gaussian error distribution characterized by a
variance $\sigma^2_{0,j}$ and the contribution $\chi^2_{j,0}$ from
the density to $\chi^2$ is then
\begin{equation}
  \chi_{j,0}^2 = [\Delta^0_j/\sigma_{0,j}]^2 = \left( \dens(\zobs_j)-\dens^\mathrm{obs}_j\right)^2/\sigma_{0,j}^2\,,
\end{equation}
where we have defined $\Delta^0_j =
\dens(\zobs_j)-\dens^\mathrm{obs}_j$, with superscript `0' to indicate
that this is a zero-th moment quantity.

The model density-weighted mean-squared velocity is given by
\begin{equation}\label{eq:modeldensv2}
  \dens\langle \vz^2\rangle(\zobs_j) = \sum_i \wi \vzi^2\,K^0(|\zobs_j+\zsun - \zi|;h)\,,
\end{equation}
where we have chosen a kernel $K_j^{\mathrm{II}}(\zi,\vzi) =
\vzi^2\,K_j^0(\zi;h)$.  As for the density, we assume that this
quantity is observed with a Gaussian error distribution with variance
$\sigma^2_{2,j}$ and the contribution $\chi^2_{j,\mathrm{II}}$ to $\chi^2$ is
\begin{equation}
  \chi_{j,\mathrm{II}}^2 = [\Delta^{\mathrm{II}}_j/\sigma_{v,j}]^2 = \left( \dens\langle v^2\rangle(\zobs_j)-\dens\langle v^2\rangle^\mathrm{obs}_j\right)^2/\sigma_{2,j}^2\,,
\end{equation}
where we have defined $\Delta^{\mathrm{II}}_j = \dens\langle
v^2\rangle(\zobs_j)-\dens\langle v^2\rangle^\mathrm{obs}_j$, where the
superscript indicates that this is a second-moment quantity (not a
square), like for the density difference $\Delta^0_j$.  The reason
that we work with the density-weighted mean-squared velocity is that
it has a simple form; for applications to data, one might want to use
the mean-squared velocity directly (for example in the example
application in \sectionname~\ref{sec:gaia} below), but this requires
normalizing by the density and thus leads to more complicated
derivatives below (see \appendixname~\ref{sec:v2}).

\begin{figure*}
  \includegraphics[width=0.99\textwidth,clip=]{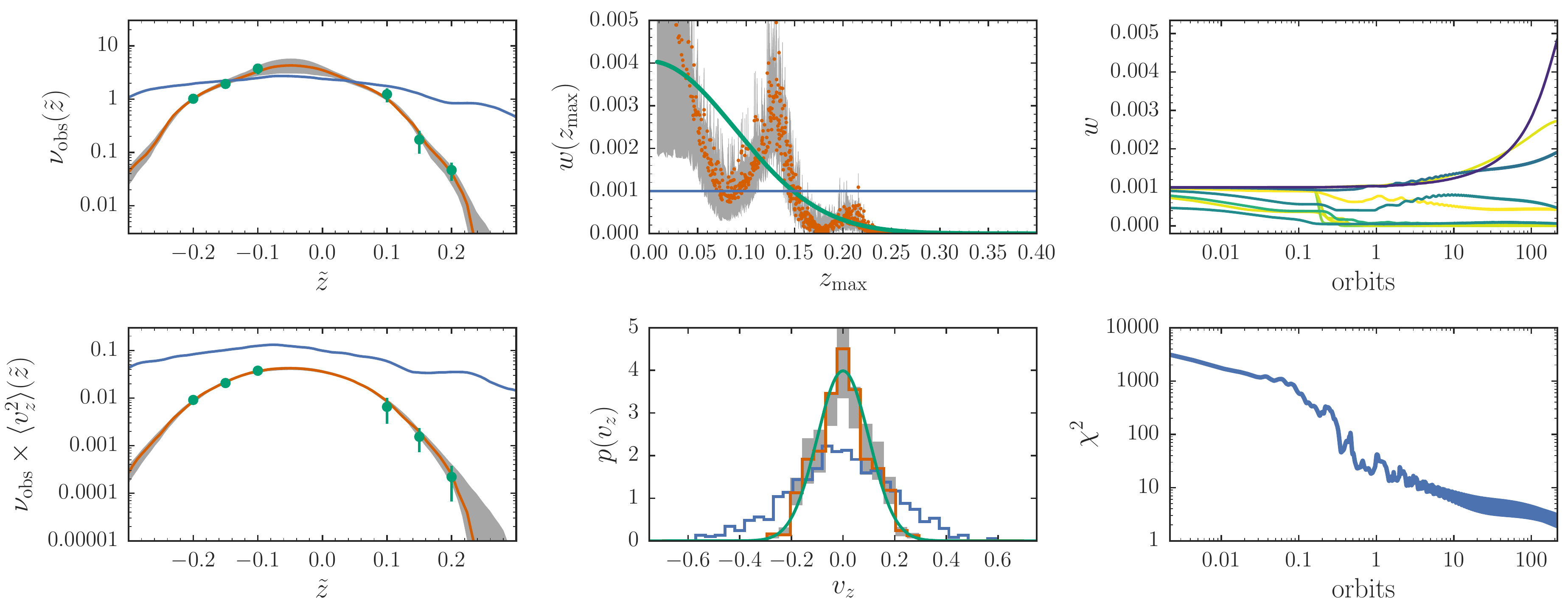}
  \caption{Basic M2M. The left panels display the observed mock data
    in green: density $\dens(\zobs)$ (top) and density-weighted
    mean-squared velocity $\dens\langle\vz^2\rangle(\zobs)$
    (bottom). The blue curve shows the initial model, while the red
    curve displays the model for the best-fit particle weights. The
    top, middle panel shows the best-fit particle weights in red, the
    initial weights in blue, and the true weights in green. The
    bottom, middle panel shows the velocity distribution (for all $z$)
    for the initial model (blue), the final model (red), and the true,
    Gaussian distribution (green). The right panels demonstrate how
    ten randomly-selected particle weights evolve (top) and how the
    total $\chi^2$ converges in the M2M optimization. The gray band in
    the left four panels displays the 68\,\% uncertainty region in the
    fit obtained from 100 samples of the PDF for the particle
    weights.}\label{fig:basic}
\end{figure*}

The standard M2M \emph{force of change} equation is then given by
\begin{align}
\label{eq:fcw}
  \frac{\dd \wi}{\dd t} & = \eps \wi\,\frac{\partial F}{\partial \wi}\\
  & = \eps \wi\left[\frac{\partial S}{\partial \wi} -\frac{1}{2}\sum_j \frac{\partial \chi^2_{j,0}}{\partial \wi}-\frac{1}{2}\sum_j \frac{\partial \chi^2_{j,\mathrm{II}}}{\partial \wi}\right]\,.\nonumber
\end{align} 
In this equation, we have that
\begin{align}
-\frac{1}{2}\frac{\partial \chi^2_{j,0}}{\partial \wi} & = -\Delta^0_j\, K^0_j(\zi;h)/\sigma_{0,j}^2\,,\\
-\frac{1}{2}\frac{\partial \chi^2_{j,\mathrm{II}}}{\partial \wi} & = -\Delta^{\mathrm{II}}_j\,\vzi^2\,K^0_j(\zi;h)/\sigma_{2,j}^2\,,
\end{align}
and
\begin{equation}
\frac{\partial S}{\partial \wi} = -\mu \ln\left[w_i/\hat{w}_i\right]\,.
\end{equation}
We solve \equationname~\eqref{eq:fcw} using a simple Euler method with
a fixed step size, computing the orbital evolution as we go along
using \equationname s~(\ref{eq:zit}) and (\ref{eq:vzit}). Unlike most
previous applications of M2M, we do not require $\sum_i \wi = 1$, but
instead let the total weight be constrained by the data (see
discussion in \sectionname~\ref{sec:discuss_aspects} below).

The M2M method for optimizing the objective function can be thought of
as a sort of gradient ascent. Gradient-ascent optimization of an
objective function does not have a physical timescale associated with
it. However, by writing the gradient-ascent algorithm in the manner of
\equationname~(\ref{eq:fcw}), we are essentially performing gradient
ascent on a clock that runs with time $\tau = \eps\,t$ compared to the
orbit integration that runs with time $t$. If $\Delta t \approx 1$ is
the orbital time scale, substantial changes to the objective function
and the particle weights only happen on timescales $1/\eps$. M2M works
by adjusting \eps\ such that $1/\eps \gg 1$, the orbital timescale,
which pushes the particle weights to an equilibrium distribution.

\subsection{Previous extensions to the standard M2M algorithm}

For the sake of completeness, we discuss some of the previous
extensions to the standard M2M method that have been proposed. These
are all concerned with how the M2M optimization for the particle
weights is run and are thus different from the extensions that we
propose in the following sections on how to fit additional parameters
beside the particle weights and how to sample from the uncertainty
distribution of all parameters.

\citet{Syer96a} propose to lessen the impact of Poisson noise due to
the finite number of $N$-body particles by smoothing the $\Delta^0_j$
and $\Delta^{\mathrm{II}}_j$ deviations that appear in
\equationname~(\ref{eq:fcw}) with smoothed versions
$\tilde{\Delta}^0_j$ and $\tilde{\Delta}^{\mathrm{II}}_j$. In the end, this leads
one to solve for ($\tilde{\Delta}^0_j$,$\tilde{\Delta}^{\mathrm{II}}_j$) using the
differential equations
\begin{align}
  \frac{\dd \tilde{\Delta}^0_j}{\dd t} & = \alpha \left(\Delta^0_j-\tilde{\Delta}^0_j\right)\,,\\
  \frac{\dd \tilde{\Delta}^{\mathrm{II}}_j}{\dd t} & = \alpha \left(\Delta^{\mathrm{II}}_j-\tilde{\Delta}^{\mathrm{II}}_j\right)\,,
\end{align}
where $\alpha$ is another inverse-timescale parameter. Because we only
want to smooth on shorter timescales than that over which we
substantially change the particle weights, we typically need $\alpha
\gtrsim \eps$ (see \citealt{Syer96a} for a detailed discussion of this
constraint). \citet{Dehnen09a} considers a modified version of this
procedure in which not the constraint but the objective function
itself is smoothed. This leads one to smooth the force-of-change
factor $\partial F / \partial \wi$ itself in a similar manner as the
\citet{Syer96a} smoothing
\begin{equation}
  \frac{\dd}{\dd t}\left(\widetilde{\frac{\partial F}{\partial \wi}} \right)= \alpha \left(\frac{\partial F}{\partial \wi} - \widetilde{\frac{\partial F}{\partial \wi}}\right)\,.
\end{equation}
Note that if we discretize the solution of
\equationname~\eqref{eq:fcw} with a stepsize $\delta t$, setting
$\alpha = 1/\delta t$ is equivalent to no smoothing and $\alpha$
cannot be set to a larger value. \citet{Malvido15a} argue that for
large particle numbers, smoothing is redundant in that the unsmoothed
algorithm already leads to final particle weights based on the
smoothed objective function. We do not apply any smoothing in any of
the examples in this paper.

\citet{Dehnen09a} also introduced a method for solving the M2M
optimization where each particle gets integrated on its own
(approximate) timescale. This is a necessary addition when modeling
systems with a wide range of orbital timescales
\citep[\eg,][]{Hunt13a} and all of our extensions of the traditional
M2M algorithm below apply to this formalism from \citet{Dehnen09a} as
well. However, we do not consider it here further, because all orbits
in our example problem of the HO have the exact same orbital
frequency.

\subsection{An example M2M fit}\label{sec:example}

\figurename~\ref{fig:basic} shows an example of the standard M2M
algorithm. We draw 100,000 mock data points from an isothermal DF with
$\sigma = 0.1$ and $\sum_i \wi = 1$ in a HO potential with $\omega =
1.3$. We evaluate the density $\dens(\zobs)$ and the density-weighted
mean-squared velocity $\dens\langle \vz^2\rangle(\zobs)$ at $\zobs =
\{\pm0.10,\pm0.15,\pm0.20\}$ for $\zsun = 0.05$ using the expressions
in \equationname s~\eqref{eq:modeldens} and \eqref{eq:modeldensv2}
with a kernel width of $h=0.025$ for an Epanechnikov kernel
\begin{equation}
  K^0(x;h) = \begin{cases}
\frac{3}{4\,h}\left[1-(\frac{x}{h})^2\right] &,\ 0 \leq x \leq h\,,\\
0&,\ \mathrm{otherwise}\,.
\end{cases}
\end{equation}
We then assume Gaussian uncertainties $\sigma_{0,j}$ and
$\sigma_{2,j}$ and obtain the measurements $\dens^\mathrm{obs}_j$ and
$\dens\langle v^2\rangle^\mathrm{obs}_j$ displayed in the left panels
of \figurename~\ref{fig:basic}. These are the measurements that we use
for all of the tests in this paper.

To model these mock data, we draw 1,000 M2M particles from the
isothermal DF with $\sigma = 0.2$---twice the true $\sigma$---and
assign them initial weights $w_i = 1/1,000$. We fix $\zsun$ and
$\omega$ to their true values. We run the standard M2M optimization
algorithm with $\eps = 10^{-3.5}$ and solve the M2M evolution with a
stepsize of $\pi/3\times 10^{-2}$ for $10^5$ steps or about 217
orbits. We do not apply a roughness penalty ($\mu = 0$) to let the
data fully determine the particle weights. We compute observables from
these 1,000 particles using a kernel with size $h = 0.075$, three
times larger than the kernel used to generate the mock data. We chose
this larger kernel to demonstrate that the kernel size or even its
shape may be different between the data and the model observables, as
long as they consistently measure the observable in question. In
\sectionname~\ref{sec:gaia}, we apply the new methods developed in
this paper to \emph{Gaia} data, where to account for the \emph{Gaia}
selection function the kernel used must be a set of rectangular bins,
while the model observables are computed using an Epanechnikov kernel,
because rectangular bins do not have well-behaved derivatives.

The resulting fit is shown in red in \figurename~\ref{fig:basic}. In
the left panels the red line is the model's density and
density-weighted mean-squared velocity evaluated at the final snapshot
of the particles with their best-fit weights. The model is smooth and
fits the data well. The top, middle panel displays the best-fit
weights $\wi$. These oscillate around their true value, indicated by
the green curve. The bottom, middle panel shows the velocity
distribution (for all $z$) of the final particle distribution. This
velocity distribution is close to a Gaussian with $\sigma = 0.1$, the
true distribution displayed in green. The right panels demonstrate how
the particle weights (top) and $\chi^2$ (bottom) converge. At the end
of the procedure we have that $\chi^2 \approx 2.5$ and we do not
optimize further (the true minimum is $\chi^2 \approx 2$). Some of the
weights that are largely unconstrained by the data are still evolving
somewhat, without affecting the model fit.

\section{Uncertainties on the particle weights}\label{sec:wunc}

The standard M2M algorithm returns the best-fit particle weights
without any estimate of their uncertainties. Standard algorithms for
sampling from the uncertainty distribution for the particle weights,
such as MCMC methods of various sorts, could in principle be applied
if we interpret the objective function in
\equationname~\eqref{eq:objective} as the logarithm of a posterior
PDF. However, these algorithms do not work well for the M2M problem,
because this posterior PDF evaluated at any given snapshot is noisy,
the weights-space is high-dimensional (dimension 1,000 in the test
example employed in this paper), and the uncertainties of the particle
weights are highly correlated.

The method for obtaining uncertainties on the particle weights that we
propose here is based on the following simple observation. Consider a
linear model in which the vector of observations $\vec{Y}$ is modeled
as a function of a parameter vector $\vec{W} \in \mathbb{R}^N$ as
$\vec{Y} = \vec{K}\,\vec{W}+\bm{\delta}$, where $\bm{\delta} \sim
\mathcal{N}(\vec{0},\vec{S})$ is Gaussian noise with mean $\vec{0}$
and known variance $\vec{S}$ (which may include correlations between
different components of $\vec{Y}$), and $\vec{K}$ is a constant
matrix. For this model, the posterior probability distribution
function (PDF) under a uniform prior is given by
\begin{equation}
  p(\vec{W} | \vec{Y},\vec{S}) = \mathcal{N}\left(\vec{M} = \vec{V}\,[\vec{K}^T\,\vec{S}^{-1}\,\vec{Y}],\vec{V}\right)\,,
\end{equation}
where the variance $\vec{V}$ is given by
\begin{equation}
  \vec{V} = [\vec{K}^T\,\vec{S}^{-1}\,\vec{K}]^{-1}\,
\end{equation}
\citep[\eg,][]{Hogg10a}. Rather than computing the mean and variance
of this Gaussian posterior PDF, we can sample from the posterior PDF
as follows
\begin{align}\label{eq:dataresampling}
  \tilde{\vec{Y}} & \sim \mathcal{N}\left(\vec{Y},\vec{S}\right)\,\\
  \tilde{\vec{M}} & = \vec{V}\,[\vec{K}^T\,\vec{S}^{-1}\,\tilde{\vec{Y}}]\,.
\end{align}
That is, we sample new observations $\tilde{\vec{Y}}$ from the
uncertainty distribution of $\vec{Y}$ and compute the `best-fit'
$\tilde{\vec{M}}$ for this new set of observations. This
$\tilde{\vec{M}}$ is a sample from the posterior PDF: (a) the
distribution of $\tilde{\vec{M}}$ is Gaussian, because
$\tilde{\vec{M}}$ is a linear transformation of another Gaussian
variable $\tilde{\vec{Y}}$, (b) the expectation value of
$\tilde{\vec{M}} = \vec{M}$, and (c) the variance $\langle
\tilde{\vec{M}}\tilde{\vec{M}}^T\rangle = \vec{V}$; because a Gaussian
distribution is fully characterized by its mean and variance, this
proves that the distribution of $\tilde{\vec{M}}$ is the correct
posterior PDF.

\begin{algorithm}[t]
    \caption{Particle Weights Monte Carlo Sampling}\label{alg:dataresampling}
    \tcc{To draw $K$ sets of particle weights $\{\wi\}_k$ for data points $\vec{Y}$ with uncertainty covariance $\vec{S}$}
    \For{$k=1,2, \ldots, K $}{
      $\tilde{\vec{Y}} \sim \mathcal{N}\left(\vec{Y},\vec{S}\right)$\\
      $\{\wi\}_k \hookleftarrow \mbox{M2M optimize}\ \wi\ \mbox{for data points}\ \tilde{\vec{Y}}$\\\qquad \qquad \ with uncertainty covariance $\vec{S}$\\
      $(\zi,\vzi) \leftarrow$ value at the end of M2M\\
      \qquad \qquad \quad \ optimization\\}
\end{algorithm}

In the M2M objective function in \equationname~\eqref{eq:objective},
the observations $\vec{Y} = Y_j$ are linearly related to the weight
parameters $\vec{W} = \wi$ through the kernels $\vec{K} =
K_j(\zi,\vzi)$.  The algorithm above is based on the assumption that
there is no constraint on the sign of each of the weight parameters
$\wi$.  Our M2M application, however, requires that all weights be non
negative for the DF to be everywhere non-negative.  To deal with this,
we therefore force the $\tilde{w}_i$ to remain positive, which is
automatically the case when using the M2M optimization algorithm
described above; we discuss the effect of this constraint in more
detail below. Thus, we sample particle weights from the weights PDF by
(a) sampling new observations $\tilde{Y}_j$ from the uncertainty
distribution for each $Y_j$, and (b) computing the best-fit particle
weights $\tilde{w}_i$ using the standard M2M algorithm (which includes
the $\tilde{w}_i> 0$ constraint) . Each such set $\tilde{w}_i$ is an
independent sample from the weights PDF, unlike in a Markov chain. We
will refer to this as the `data-resampling method for sampling the
particle weights PDF'. This method is presented in
Algorithm~\ref{alg:dataresampling}. The algorithm, as written down
there, draws $K$ samples from the uncertainty distribution for the
particle weights; when we use this algorithm as part of a larger Gibbs
MCMC chain, we will typically use it to draw just a single sample
($K=1$ in Algorithm~\ref{alg:dataresampling}).

This method does not properly deal with particle weights for which the
penalty term in \equationname~\eqref{eq:entropy} (which becomes a
prior when sampling the particle weights) has a significant effect or
for weights that, if they were allowed to be negative, have much
probability mass at $\wi < 0$. An extreme case of the former are
weights of orbits that do not pass through any observed volume. Under
the optimization algorithm, these will always return the prior weight
with no scatter. If the prior on the particle weights was Gaussian we
could similarly sample new prior means as the first step in the
algorithm in \equationname~\eqref{eq:dataresampling} (because the
prior mean $\hat{w}_i$ is in this case equivalent to an `observation'
of $\wi$ with an error variance equal to the prior variance). We do
not implement this here, but see further discussion of this in
\sectionname~\ref{sec:discuss_future}. For weights that want $\wi <
0$, the optimization algorithm will effectively associate all
probability mass at $\wi < 0$ with $\wi = 0$. While this is not
technically correct---it does not sample from the posterior PDF---it
is reasonable to set weights to zero that want to be less than
zero. Some M2M algorithms remove orbits with small or zero weights and
our sampling method effectively samples from the two alternative
models for such orbits with the probability of these two alternatives
determined by the data: (a) they get removed ($\wi = 0$) and (b) they
have non-zero weights ($\wi > 0$). We discuss the issue of particle
weights that prefer to be negative in the context of the M2M
extensions in the next two sections further in
\sectionname~\ref{sec:discuss_weights}.

\begin{figure*}
  \includegraphics[width=0.99\textwidth,clip=]{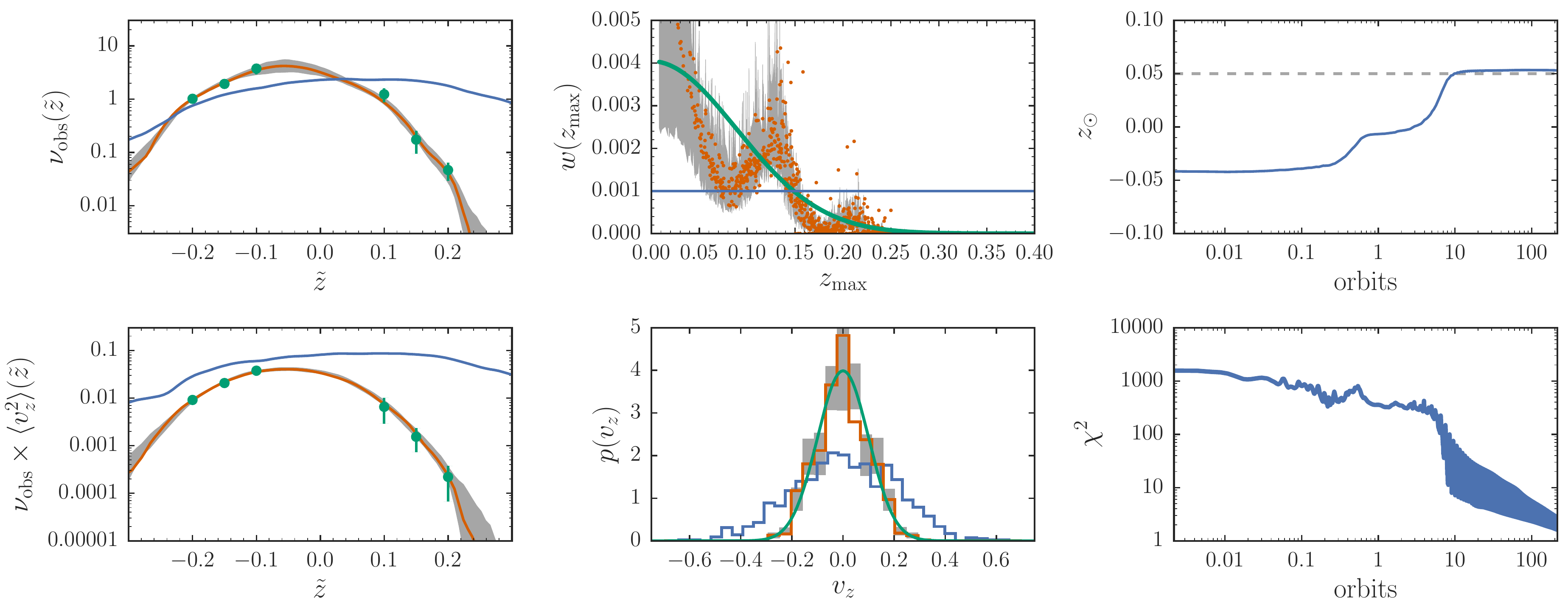}
  \caption{M2M with nuisance parameters. Like
    \figurename~\ref{fig:basic}, except that we also fit for the Sun's
    height above the plane \zsun\ using the force of change for
    \zsun\ and we sample the uncertainty in both the particle weights
    and \zsun. The top, right panel demonstrates how \zsun\ converges
    during the joint M2M optimization of the particle weights and
    \zsun. We find that $\zsun = 0.0527\pm0.0042$, in good agreement
    with its true value of 0.0500, shown as the dashed gray line in
    the top, right panel.}\label{fig:zsun}
\end{figure*}

An example of the data-resampling method for sampling the
particle-weights PDF is shown in \figurename~\ref{fig:basic}. We draw
100 samples from the weights PDF, that is, 100 sets of 1,000 particle
weights. Each set is optimized using the same optimization settings as
in \sectionname~\ref{sec:example}; each set's initial particle
distribution is set to the final snapshot of the previous
sample. Because we set $\mu=0$, we assume a flat, improper prior $\wi
> 0$ for all particle weights. The gray band displays the
$\approx1\sigma$ range spanned by this sample of particle weights. The
uncertainty in the particle weights (top, middle panel) and consequent
uncertainty in the density and density-weighted mean-squared velocity
(left panels) and the velocity distribution (bottom, middle panel)
adheres to our physical intuition. For example, orbits with
$z_\mathrm{max} < 0.05$ are essentially only constrained by the
observations at $\zobs = -0.1$, which corresponds to $z = -0.05$
because $\zsun = 0.05$; the uncertainty in the particle weights blows
up at $z_\mathrm{max} < 0.05$ because of this. The density kernel for
an observation at $z$ is dominated by orbits with $z_\mathrm{max}
\approx z$, while the velocity-squared kernel at all $z$ gets large
contributions from stars with large $z_\mathrm{max}$. Therefore,
weights at high $z_\mathrm{max}$ are strongly constrained by the
velocity data. The uncertainty in the density in the left panel is
therefore large near $\zobs \approx 0$, while the uncertainty in the
velocity is small at the same $\zobs$. At large $\zobs$ the data allow
a more steeply declining density and/or velocity, but not a shallower
distribution (which would have too large velocities at low
heights). Keep in mind that these strong relations depend on knowing
the gravitational potential and keeping it fixed.

\section{Optimizing and sampling nuisance parameters}\label{sec:nuisance}

\begin{algorithm}[t]
    \caption{MCMC sampling of nuisance parameters}\label{alg:nuisance}
    \tcc{To draw $K$ MCMC samples $z_{\odot,k}$, given a set of particle weights $\{\wi\}$ and a gravitational potential, for data points $\vec{Y}$ with uncertainty covariance $\vec{S}$}
    \tcp{Average objective function for current \zsun:}
    $\tilde{F} \leftarrow 0$\\
    \For{$m=1,2, \ldots, M $}{
      $(\zi,\vzi) \leftarrow$ advance orbits by 1 step\\
      $\tilde{F}+= F(\zi,\vzi|\zsun,\wi,\vec{Y},\vec{S})/M$\\
    }
    \tcp{MCMC sample using Metropolis-Hastings:}
    \For{$k=1,2, \ldots, K $}{
      \tcp{Draw proposed $\zsun'$:}
        $\zsun' \sim Q(\zsun'|\zsun)$\\
      $(\zi,\vzi) \leftarrow$ rewind orbits by $M$ steps\\
      \tcp{Average objective function for $\zsun'$:}
      $\tilde{F}' \leftarrow 0$\\
        \For{$m=1,2, \ldots, M $}{
          $(\zi,\vzi) \leftarrow$ advance orbits by 1 step\\
          $\tilde{F}'+= F(\zi,\vzi|\zsun',\wi,\vec{Y},\vec{S})/M$\\
        }
        \tcp{Accept/reject:}
        $q \leftarrow \tilde{F}'-\tilde{F}$\\
        $r \sim [0,1]$\\
        \eIf{$\ln r < q$}{
          $\zsun \leftarrow \zsun'$\\
          $\tilde{F} \leftarrow \tilde{F}'$\\
        }{
          $\zsun \leftarrow \zsun$}
        $z_{\odot,k} \leftarrow \zsun$\\
    }
\end{algorithm}

\begin{figure}
  \includegraphics[width=0.44\textwidth,clip=]{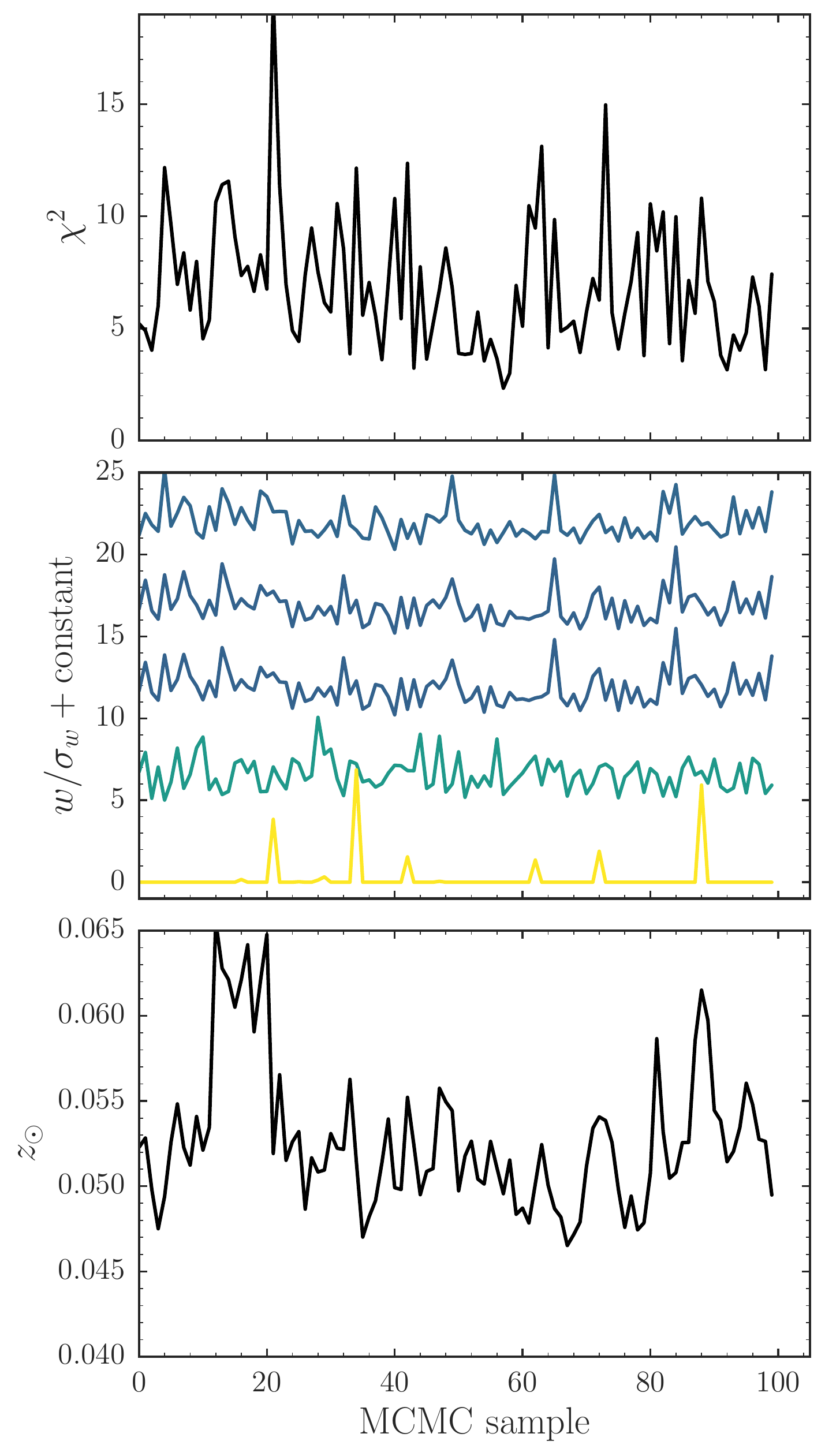}
  \caption{MCMC sampling of the particle weights and \zsun. This
    figure demonstrates the MCMC chain of 100 samples from the
    uncertainty distribution of the particle weights and
    \zsun\ constrained by the mock data. The top panel displays the
    behavior of $\chi^2$, the middle panel that of 5 random particle
    weights (normalized by the standard deviation of their samples,
    color-coded by $z_{\mathrm{max}}$), and the bottom panel shows the
    \zsun\ samples. The chain has a small correlation length, because
    we perform 20 Metropolis-Hastings steps for \zsun\ for each sample
    from the weights PDF.}\label{fig:zsun_mcmc}
\end{figure}

Dynamical modeling of observed galaxy kinematics often requires the
knowledge of parameters separate from those specifying the
distribution function (the particle weights in the M2M case) and those
related to the gravitational potential. These are typically related to
the observer's perspective: for example, the observer's
three-dimensional position and velocity with respect to the center of
the system being modeled (\eg, the Sun's distance from the Galactic
center for Milky-Way dynamics) or the observer's viewing angle (\eg, a
galaxy's inclination for an external galaxy, the Sun's position with
respect to the bar when modeling the central Milky Way). These types
of parameters enter into the kernel evaluation in the M2M objective
function. The standard method for determining these parameters is to
optimize the M2M objective function for the particle weights on a grid
of nuisance parameters. Here we demonstrate that the M2M objective
function in \equationname~\eqref{eq:objective} can be optimized
simultaneously for the particle weights and the nuisance parameters.

{\allowdisplaybreaks As an example we consider the Sun's height
  $\zsun$ above the plane. The Sun's height enters the kernels through
  $z = \zobs + \zsun$. Similar to the standard M2M algorithm, we can
  form a force of change equation for \zsun\ as
\begin{align}\label{eq:fcz}
  \frac{\dd \zsun}{\dd t} & = \eps_\odot \,\frac{\partial F}{\partial \zsun}\\
  & = \eps_\odot \left[-\frac{1}{2}\sum_j \frac{\partial \chi^2_{j,0}}{\partial \zsun}-\frac{1}{2}\sum_j \frac{\partial \chi^2_{j,\mathrm{II}}}{\partial \zsun}\right]\,.\nonumber
\end{align}
where we have allowed the freedom to use a different $\eps_\odot$ from
the $\eps$ parameter used in the force-of-change equation for the
particle weights. We have that
\begin{align}
-\frac{1}{2}\, \frac{\partial \chi^2_{j,0}}{\partial \zsun} & = - \frac{\Delta^0_j}{\sigma_{0,j}^2}\,\mathlarger{\mathlarger{\sum}}_{i}\wi \frac{\dd K^0_j(r;h)}{\dd r}\Bigg|_{|\zobs_j+\zsun-\zi|}\!\!\!\!\!\!\!\!\!\!\!\!\!\!\!\!\!\mathrm{sign}(\zobs_j+\zsun-\zi)\,,\\
-\frac{1}{2}\, \frac{\partial \chi^2_{j,\mathrm{II}}}{\partial \zsun} & = - \frac{\Delta^{\mathrm{II}}_j}{\sigma_{2,j}^2}\,\mathlarger{\mathlarger{\sum}}_{i} \wi\,\vzi^2 \frac{\dd K^0_j(r;h)}{\dd r}\Bigg|_{|\zobs_j+\zsun-\zi|}\!\!\!\!\!\!\!\!\!\!\!\!\!\!\!\!\!\!\!\!\!\!\!\!\mathrm{sign}(\zobs_j+\zsun-\zi)\,.
\end{align}
If one wants to include a prior on \zsun\ there would be an additional
contribution to the force of change from this prior. We then again
solve the system of \equationname s~\eqref{eq:fcw} and \eqref{eq:fcz}
using an Euler method with a fixed step size, computing the orbital
evolution as we go along using \equationname s~(\ref{eq:zit}) and
(\ref{eq:vzit}).}

An example of this is displayed in \figurename~\ref{fig:zsun}, where
we fit the same data as in the example described in
\sectionname~\ref{sec:example}, but now also fitting \zsun. All of the
optimization parameters are kept the same and we set $\eps_\odot =
10^{-6} \approx \eps/300$. We start at an initial guess of $\zsun =
-0.05$, far from the true value. We see that $\zsun$ quickly and
smoothly converges to $\zsun = 0.053$, close to the true value.

After finding the best-fit \zsun\ from the M2M optimization, we can
sample the joint posterior PDF for $(\wi,\zsun)$ using a
Metropolis-Hastings-within-Gibbs sampler by repeating the following
steps
\begin{align}\label{eq:mhzsun}
  (a)\quad \wi & \sim p(\wi|\zsun,\mathrm{observations})\,,\quad [\mathrm{Algorithm}~\ref{alg:dataresampling}]
  \\ (b)\quad \zsun & \sim
  p(\zsun|\wi,\mathrm{observations})\,,\quad [\mathrm{Algorithm}~\ref{alg:nuisance}]\,,\label{eq:mhzsun2}
\end{align}
where we sample particle weights in the (a) step using the
data-resampling technique of \sectionname~\ref{sec:wunc} (see
Algorithm~\ref{alg:dataresampling} with $K=1$ to draw a single
particle-weights sample) and sample $\zsun$ using a
Metropolis-Hastings (MH) update using the objective function as the
log posterior PDF $\ln p(\zsun|\wi,\mathrm{observations})$, in which
the weights $\wi$ are held fixed. Step (b) is presented in detail in
Algorithm~\ref{alg:nuisance}. In practice, we average the objective
function in step (b) over about one orbital period (lines 2--5 and
10--13 in Algorithm~\ref{alg:nuisance}) and use the exact same orbital
trajectories (thus the rewind step in line 8 of
Algorithm~\ref{alg:nuisance}) to reduce the noise in the objective
function. Because the optimization in step (a) typically requires tens
to hundreds of orbital periods, step (b) proceeds quickly compared to
step (a). We can improve mixing in the MCMC chain by performing
multiple MH steps for each weights sample ($K > 1$ in
Algorithm~\ref{alg:nuisance}) and keeping only the final \zsun\ sample
in each step (b); as long as the total number of orbital steps in (b)
is much less than that for a single optimization, this does not
increase the computational cost significantly.

The result of this procedure for the example problem is shown in
\figurename s~\ref{fig:zsun} and \ref{fig:zsun_mcmc}. We have drawn
100 samples from the joint PDF of the particle weights and \zsun,
using a Gaussian proposal distribution with standard deviation
$\sigma_{\zsun} = 0.01$ and performing $10^5$ M2M optimization time
steps in step (a) and 20 MH steps for each particle-weights
sample. The chain is initialized at the best-fit $\zsun$ from the M2M
optimization described above. We average the objective function using
$M=500$ steps or about 1 orbital period. The behavior of the MCMC
chain is displayed in \figurename~\ref{fig:zsun_mcmc}. This figure
demonstrates that the chain is well-mixed and has a small correlation
length (adjacent samples have very different values). The chain for
the particle weights demonstrates that weights with similar
$z_\mathrm{max}$ are strongly correlated. The acceptance ratio for the
Metropolis-Hastings steps for \zsun\ for this chain is 0.30.

The uncertainty in the density and velocity profiles in
\figurename~\ref{fig:zsun} now includes the uncertainty in \zsun\ and
this increases the overall uncertainty. We find that $\zsun =
0.0527\pm0.0042$. We can compare this to the standard method of
constraining \zsun: we optimize the particle weights for a set of
fixed \zsun\ and record the minimum $\chi^2$ for each $\zsun$. This
gives $\zsun = 0.0534\pm0.0046$. We can also compare our M2M-based
result to the result if we assume that the DF is isothermal with
unknown $\sigma$ and normalization. In that case, the data constrain
$\zsun = 0.0560\pm0.0048$, similar to the M2M analyses. Overall, we
find that the novel M2M procedure performs well.

\section{Optimizing and sampling the gravitational potential}\label{sec:potential}

\begin{figure*}
  \includegraphics[width=0.99\textwidth,clip=]{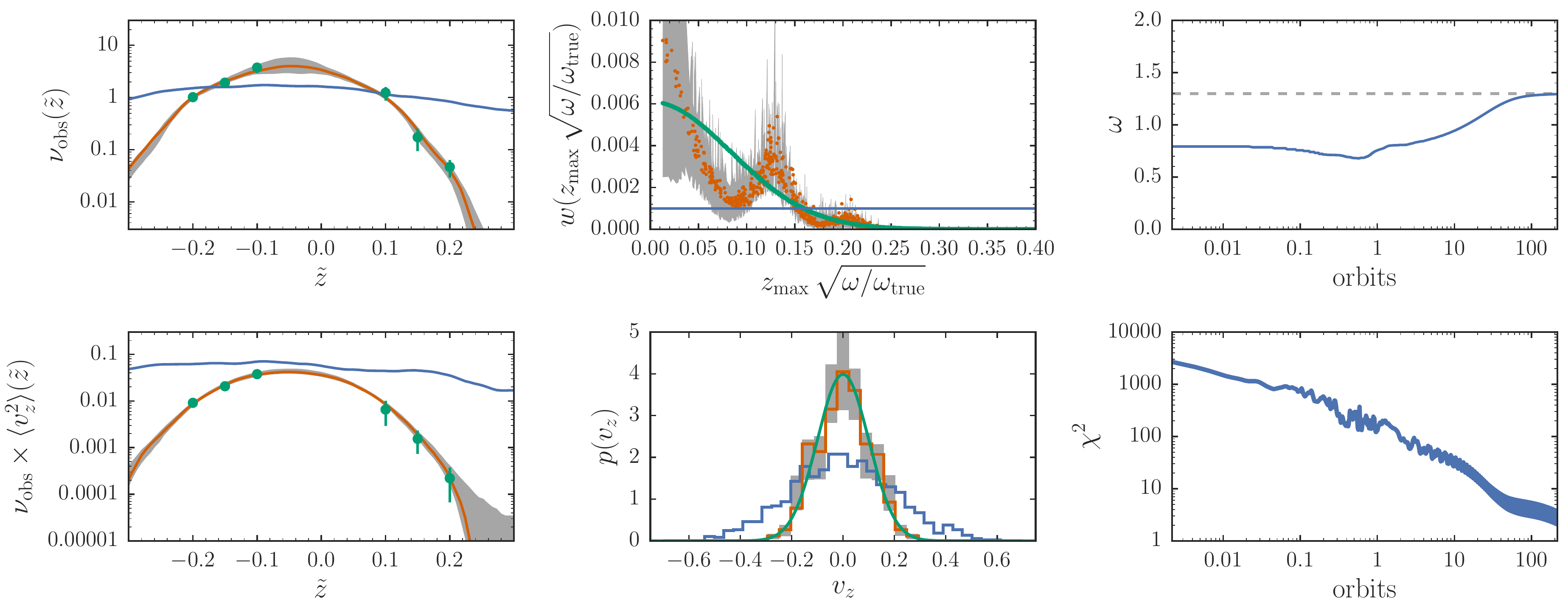}
  \caption{M2M for the parameters of the gravitational potential. Like
    \figurename~\ref{fig:basic}, except that we also fit for the HO
    potential's frequency $\omega$ using the force of change for
    $\omega$ and we sample the uncertainty in both the particle
    weights and $\omega$. Because $z_\mathrm{max}$ is not conserved
    when changing $\omega$, we plot the weights as a function of
    $z_\mathrm{max}\sqrt{\omega}$, which is proportional to the square
    root of the action, which is approximately conserved during the
    M2M fit and sampling. The top, right panel demonstrates how
    $\omega$ converges during the joint M2M optimization of the
    particle weights and $\omega$. We find that $\omega =
    1.32\pm0.08$, in good agreement with its true value of 1.3,
    indicated by the dashed gray line in the top, right
    panel.}\label{fig:omega}
\end{figure*}

\begin{algorithm}[t]
    \caption{MCMC sampling of potential parameters}\label{alg:potential}
    \tcc{To draw $K$ MCMC samples $\omega_k$ characterizing potentials $\Phi(z;\omega_k)$, given a set of particle weights $\{\wi\}$ and nuisance parameter $\zsun$, for data points $\vec{Y}$ with uncertainty covariance $\vec{S}$}
    \tcp{Average objective function for current $\omega$:}
    $\tilde{F} \leftarrow 0$\\
    \For{$m=1,2, \ldots, M $}{
      $(\zi,\vzi) \leftarrow$ advance orbits by 1 step in $\Phi(z;\omega)$\\
      $\tilde{F}+= F(\zi,\vzi|\omega,\zsun,\wi,\vec{Y},\vec{S})/M$\\
    }
    \tcp{MCMC sample using Metropolis-Hastings:}
    \For{$k=1,2, \ldots, K $}{
      \tcp{Draw proposed $\omega'$:}
      $\omega' \sim Q(\omega'|\omega)$\\
      \tcp{Adiabatically change $\omega$ to $\omega'$:}
      $(\zi',\vzi') \leftarrow (\zi,\vzi)$\\
      \For{$l=1,2, \ldots, L $}{
        $\omega_l \leftarrow \omega + (\omega'-\omega)\,l/L$\\
        $(\zi',\vzi') \leftarrow$ rewind orbits by 1 step in $\Phi(z;\omega_l)$\\
      }
      $(\zi',\vzi') \leftarrow$ advance orbits by $L-M$ steps in\\
      \quad \quad \quad \quad \ \ \  $\Phi(z;\omega')$\\
      \tcp{Average objective function for $\omega'$:}
        $\tilde{F}' \leftarrow 0$\\
        \For{$m=1,2, \ldots, M $}{
          $(\zi',\vzi') \leftarrow$ advance orbits by 1 step in\\
          \quad \quad \quad \quad \quad $\Phi(z;\omega')$\\
          $\tilde{F}'+= F(\zi',\vzi'|\omega',\zsun,\wi,\vec{Y},\vec{S})/M$\\
        }
        \tcp{Accept/reject}
        $q \leftarrow \tilde{F}'-\tilde{F}$\\
        $r \sim [0,1]$\\
        \eIf{$\ln r < q$}{
          $\omega \leftarrow \omega'$\\
          $\tilde{F} \leftarrow \tilde{F}'$\\
          $(\zi,\vzi) \leftarrow (\zi',\vzi')$\\
        }{
          $\omega \leftarrow \omega$}
        $\omega_k \leftarrow \omega$\\
    }
\end{algorithm}

\begin{figure}
  \includegraphics[width=0.44\textwidth,clip=]{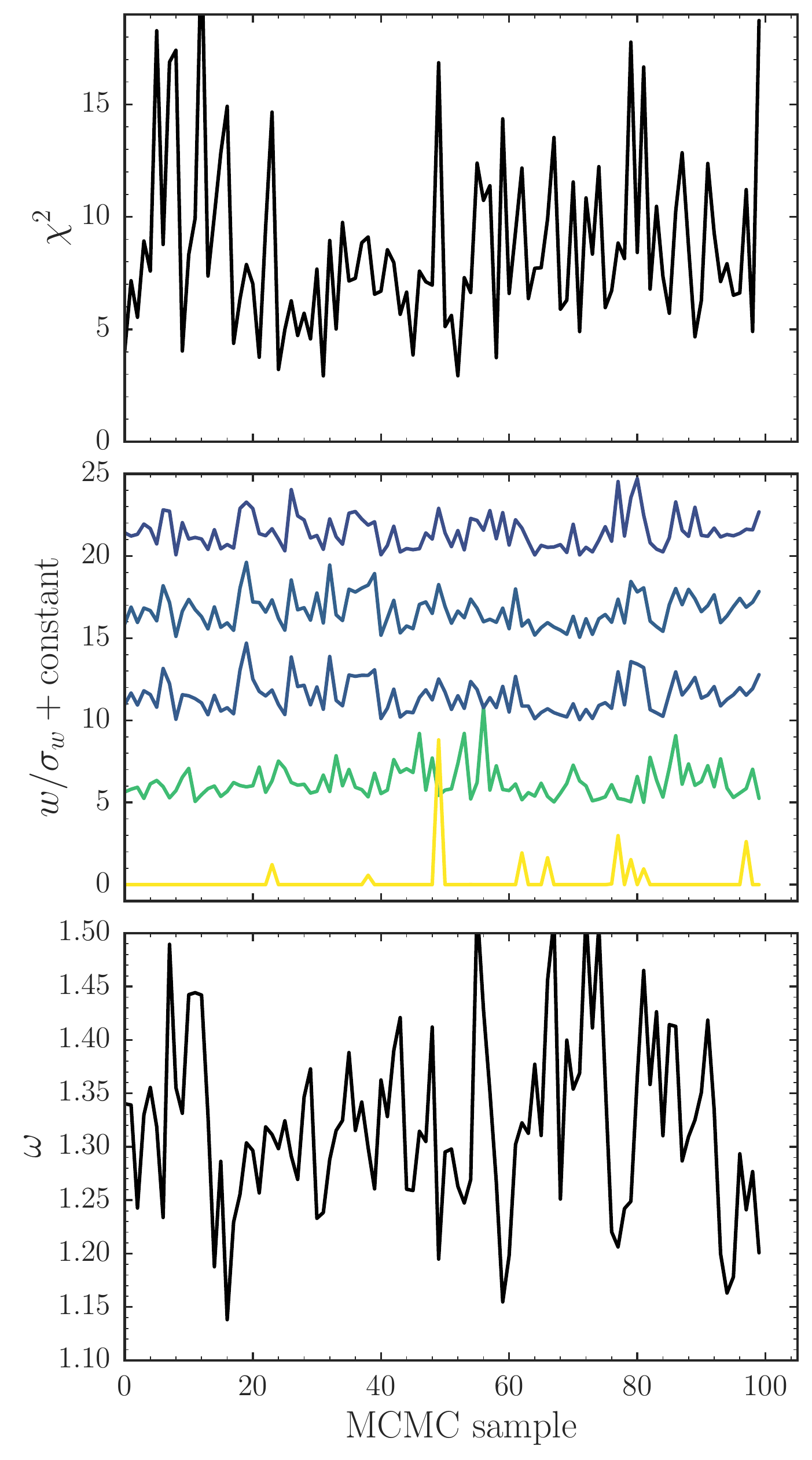}
  \caption{MCMC sampling of the particle weights and $\omega$. This
    figure demonstrates the MCMC chain of 100 samples from the
    uncertainty distribution of the particle weights and $\omega$
    constrained by the mock data. The top panel displays the behavior
    of $\chi^2$, the middle panel that of 5 random particle weights
    (normalized by the standard deviation of their samples,
    color-coded by $z_{\mathrm{max}}\sqrt{\omega}$), and the bottom
    panel shows the $\omega$ samples. The chain has a small
    correlation length, because we perform 20 Metropolis-Hastings
    steps for $\omega$ for each sample from the weights
    PDF.}\label{fig:omega_mcmc}
\end{figure}

\begin{figure*}
  \includegraphics[width=0.99\textwidth,clip=]{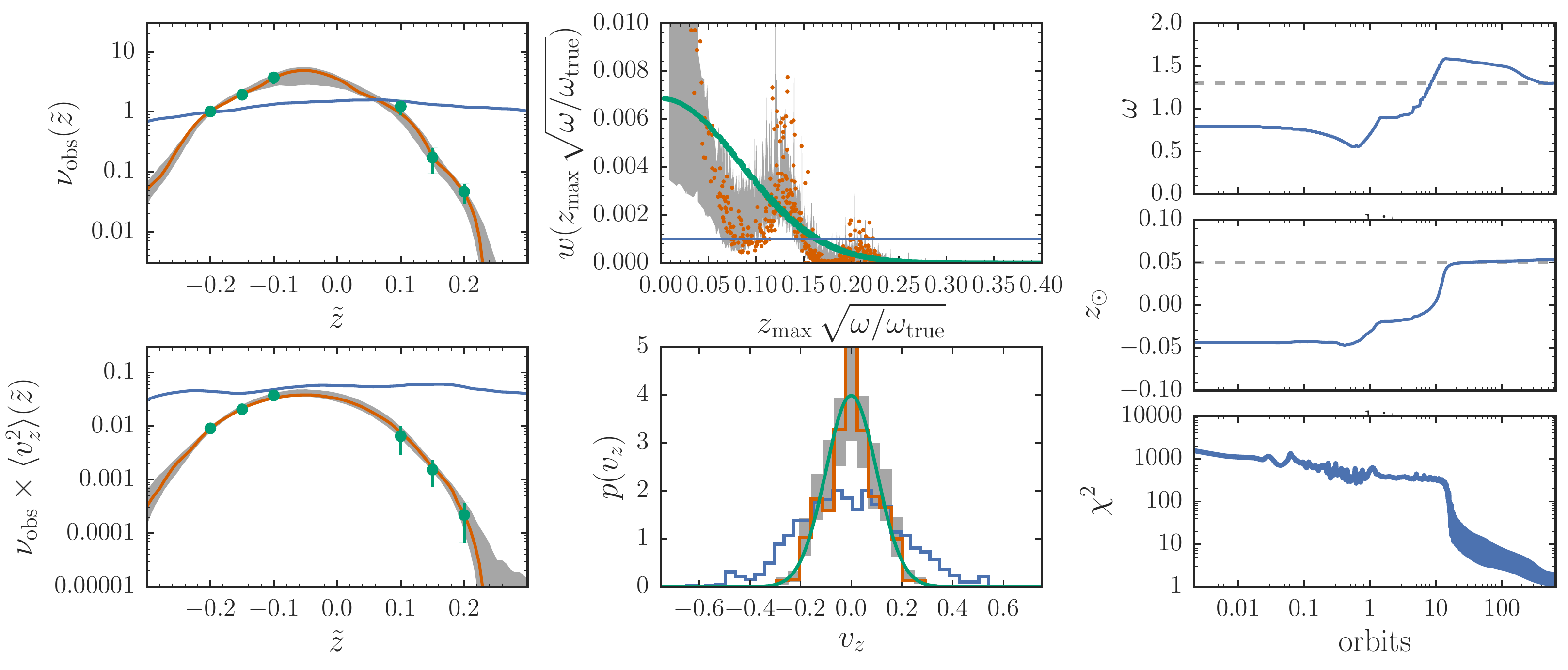}
  \caption{Full probabilistic M2M modeling. Like
    \figurename~\ref{fig:basic}, except that we also fit for the HO
    potential's frequency $\omega$ as well as the Sun's height above
    the plane $\zsun$. We MCMC sample from the joint PDF for the
    particle weights, $\zsun$, and $\omega$. As in
    \figurename~\ref{fig:omega}, because $z_\mathrm{max}$ is not
    conserved when changing $\omega$, we plot the weights as a
    function of $z_\mathrm{max}\sqrt{\omega}$, which is proportional
    to the square root of the action. The top, right and middle panels
    demonstrate how $\omega$ and $\zsun$, respectively, converge
    during the joint M2M optimization of the particle weights,
    $\omega$, and \zsun. We find that $\omega = 1.316\pm0.085$ and
    $\zsun = 0.053\pm0.005$, in good agreement with their true values
    of $\omega_{\mathrm{true}} = 1.3$ and $z_{\odot,\mathrm{true}} =
    0.05$ (dashed lines in the right panels).}\label{fig:full}
\end{figure*}

Traditionally, M2M modeling, much like Schwarzschild modeling, keeps
the external gravitational field fixed during the M2M fit. The
gravitational potential is optimized by running the fit for different
fixed potentials and choosing the potential that provides the best
fit. While the overall distance and velocity scale can be optimized by
writing down a force of change equation for these
\citep{DeLorenzi08a}, this does not apply to other parameters of the
potential. However, similar to the force of change for nuisance
parameters, we can write down the force of change for parameters
describing the potential and adjust these parameters during the
fit. Naively, the problem with this procedure is that the
instantaneous objective function $F$ does not depend on the potential,
because it is only a function of the current phase--space position of
the M2M particles. In this section we discuss how to get around this
problem, such that we can fit and MCMC sample the parameters
describing the gravitational potential.

{\allowdisplaybreaks
Using our HO example, we vary the frequency $\omega$ of the HO
potential. The force of change equation for $\omega$ is
\begin{align}\label{eq:fcomega}
  \frac{\dd \omega}{\dd t} & = \eps_\omega \,\frac{\partial F}{\partial \omega}\nonumber\\
  & = \eps_\omega \left[-\frac{1}{2}\sum_j \frac{\partial \chi^2_{j,0}}{\partial \omega}-\frac{1}{2}\sum_j \frac{\partial \chi^2_{j,\mathrm{II}}}{\partial \omega}\right]\,.\nonumber\\
  & = - \eps_\omega \left[\frac{\Delta^0_j}{\sigma_{0,j}^2}\,\frac{\partial \Delta^0_j}{\partial \omega}+\frac{\Delta^{\mathrm{II}}_j}{\sigma_{2,j}^2}\,\frac{\partial \Delta^{\mathrm{II}}_j}{\partial \omega}\right]\,.
\end{align}
where we have again allowed the freedom to use a different
$\eps_\omega$ from the $\eps$ parameter used in the force-of-change
equation for the particle weights or for the nuisance parameters. We
can directly evaluate $\frac{\partial \Delta^0_j}{\partial
  \omega}$ and $\frac{\partial \Delta^{\mathrm{II}}_j}{\partial \omega}$ using a
finite difference approximation, \eg, $\frac{\partial
  \Delta^0_j}{\partial \omega} =
\frac{\Delta^0_j(\omega+\Delta
  \omega)-\Delta^0_j(\omega)}{\Delta \omega}$, by integrating the
orbit starting from the previous time step in the two potentials
characterized by frequencies $\omega$ and $\omega + \Delta \omega$ and
comparing the $(\Delta^0_j,\Delta^{\mathrm{II}}_j)$ at the current time. In
practice, we compute these finite differences with a $\Delta \omega$
large enough to give a substantial difference in $(\zi,\vzi)$ over the
time step $\Delta t$. The parameter $\eps_\omega$ should be small
enough such that substantial changes to $\omega$ only happen on many
orbital timescales. In that case, the (non-resonant) orbits change
adiabatically and the orbital structure corresponding to the M2M
particles does not change between potentials. In certain applications
it may also be necessary to adiabatically change the potential to
that, in this case, corresponding to $\omega + \Delta \omega$ when
computing the finite difference, but we do not find this to be
necessary here.}

An example of fitting for $\omega$ is shown in
\figurename~\ref{fig:omega}, where we fit the same data as in the
example described in \sectionname~\ref{sec:example}, but now also
fitting $\omega$ (while keeping $\zsun$ fixed to its true value). We
keep the optimization parameters for the particle weights the same as
in \sectionname~\ref{sec:example}, but use $\eps_\omega = 10^{-3}$. We
compute the finite difference using \equationname~\eqref{eq:fcomega}
with $\Delta \omega = 0.3$ and we only update $\omega$ every 10 time
steps (and we therefore compute the finite difference using a time
step $\Delta t = $10 times the basic stepsize). We start at an initial
guess $\omega = 0.8$ and the fit converges to $\omega = 1.297$, close
to the true value $(\omega = 1.3$).

Like for nuisance parameters, we can sample the joint posterior PDF
for the particle weights and the potential parameters, in this case
$\omega$, using Metropolis-Hastings-within-Gibbs. The full MCMC
sampler including the nuisance parameter $\zsun$ is then given by
\begin{align}\label{eq:mhomega}
  (a)\quad \wi & \sim p(\wi|\zsun,\omega,\mathrm{observations})\,,\ [\mathrm{Algorithm}~\ref{alg:dataresampling}]\,,\\ 
  (b)\quad \zsun & \sim  p(\zsun|\omega,\wi,\mathrm{observations})\,,\ [\mathrm{Algorithm}~\ref{alg:nuisance}]\,,\label{eq:mhzsun2}\\\
  (c)\ \quad \omega & \sim  p(\omega|\zsun,\wi,\mathrm{observations})\,,\ [\mathrm{Algorithm}~\ref{alg:potential}]\,,\label{eq:mhomega2}
\end{align}
where we again sample particle weights in the (a) step using the
data-resampling technique of \sectionname~\ref{sec:wunc} (using $K=1$
to draw a single particle-weights sample) and sample $\zsun$ and
$\omega$ in steps (b) and (c) using a Metropolis-Hastings (MH) update
using the objective function as the log posterior PDF, presented in
detail in Algorithms~\ref{alg:nuisance} and \ref{alg:potential}. Like
for the nuisance parameters on their own, we average the objective
function in steps (b) and (c) over about one orbital period. Rather
than simply changing the potential abruptly from a frequency $\omega$
to a proposal $\omega'$ for the likelihood evaluation in step (c), we
adiabatically change the potential parameter from its current value to
its proposed value before evaluating the likelihood (lines 8--12 in
Algorithm~\ref{alg:potential}). We perform this adiabatic change by
integrating backwards in time and then partially integrating forwards
in time, in such a way that the subsequent likelihood evaluation would
use the exact same orbital trajectories if $\omega$ were not changed
(line 13 in Algorithm~\ref{alg:potential}). This reduces the noise
from the particle distribution in the likelihood evaluation. We can
again improve mixing in the MCMC chain by performing multiple MH steps
(b) and (c) for each particle-weights sample ($K > 1$ in
Algorithms~\ref{alg:nuisance} and \ref{alg:potential}, not necessarily
equal) and keeping only the final $\omega$ sample in each step
(c). The adiabatic change of the potential is important for
maintaining the reversibility of the MCMC chain. If the potential is
changed non-adiabatically, orbits differ when revisiting the same
potential $\Phi(z;\omega)$ and the likelihood of a given set of
particle weights is then different at later times. This does not
happen when the potential is changed adiabatically, because the nature
of the orbits represented by the M2M particles does not change.

We apply this MCMC algorithm to sample the uncertainty distribution of
the particle weights and $\omega$ given the mock data, fixing $\zsun$
to its true value [that is, skipping step (b)]. In step (c), we use a
Gaussian proposal with standard deviation $\sigma_\omega = 0.2$ and
again perform $10^5$ M2M optimization time steps in step (a) and 20 MH
steps for each step (a). We adiabatically change the potential using
$L = 10,000$ steps---or about 20 orbital periods---and average the
objective function using 1,000 orbital time steps. The MCMC chain is
started at the best-fitting $\omega$ in the M2M optimization described
above. The behavior of the MCMC chain is displayed in
\figurename~\ref{fig:omega_mcmc}. The MH acceptance ratio for the
$\omega$ steps in the chain is 0.37. The chain is again well-mixed and
has a short correlation length.

From the MCMC samples we find that the mock data constrain $\omega =
1.32\pm0.08$. We can compare this to the standard M2M procedure, where
the PDF for $\omega$ is approximated using the best-fit particle
weights for each trial $\omega$. This gives $\omega = 1.31\pm0.08$,
similar to the MCMC result. If we assume that the DF is isothermal and
marginalize over the amplitude and velocity dispersion of this
isothermal DF, we find $\omega = 1.19\pm0.07$. All of these are
consistent with the true value $\omega_{\mathrm{true}} = 1.3$. That
the isothermal DF gives a different best-fit $\omega$ than the M2M
modeling is unsurprising, because it fits a different functional shape
to the density and velocity constraints: the best-fit M2M DF is close
to, but not exactly isothermal.

As a final test problem, we fit the particle weights, nuisance
parameter $\zsun$, and the potential parameter $\omega$ simultaneously
to the mock data and then perform full MCMC sampling using steps (a)
through (c) above. For the optimization part, we use
$(\eps,\eps_{\odot},\eps_\omega) = (10^{-3.5},10^{-6},10^{-3})$ and
integrate for $3\times10^5$ time steps, again updating $\omega$ only
every 10 time steps. Otherwise the setup is the same as above. We use
the best-fit $(\zsun,\omega)$ as the initial condition for MCMC
sampling. In the MCMC sampling, we use $10^5$ optimization time steps
in step (a) of the algorithm and we average the likelihood using 500
steps for sampling $\zsun$ and using 1,000 steps for sampling $\omega$
and again adiabatically change the frequency in MH steps over 10,000
time steps. The result is shown in \figurename~\ref{fig:full}. The
parameters \zsun\ and $\omega$ converge to best-fit values of $\zsun =
0.0530$ and $\omega = 1.27$. From the MCMC chain we find that $\zsun =
0.053\pm0.005$ and $\omega = 1.316\pm0.085$, similar to the analyses
where one of these was kept fixed at its true value.

\section{Application to \emph{Gaia} DR1}\label{sec:gaia}

As a first real-data application of the M2M extensions described in
this paper, we model the vertical dynamics of F-type dwarfs using data
from the \emph{Gaia} DR1 \emph{Tycho-Gaia Astrometric Solution}
(\emph{TGAS}; \citealt{GaiaDR1,Lindegren16a}). We stress that the
point of this application is only to illustrate the performance of the
new M2M method on real data. Because we use the same HO model for the
potential, which is not a fully realistic model for the vertical
potential near the Sun, the parameter constraints that we derive below
cannot be easily translated into a constraint on the local mass
distribution and we do not attempt to put any constraint on the local
gravitational potential from this modeling.

\begin{figure*}
  \includegraphics[width=0.99\textwidth,clip=]{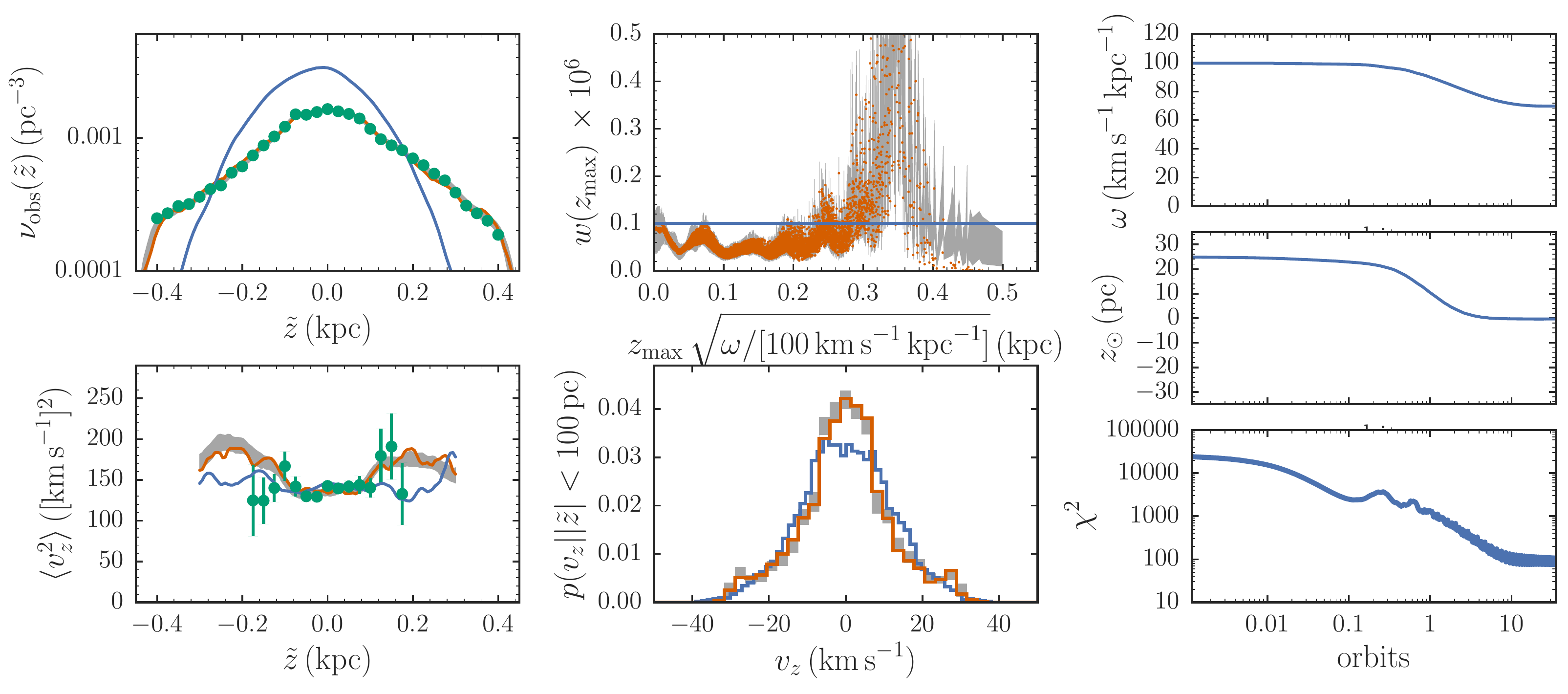}
  \caption{Probabilistic M2M modeling of the vertical dynamics of
    F-type dwarfs in \emph{Gaia} DR1. Panels are the same as in
    \figurename~\ref{fig:full}, except that we model the mean-squared
    velocity directly rather than the density-weighted mean-squared
    velocity (bottom left panel) and we display the local velocity
    distribution in the bottom middle panel. We can successfully model
    the density and the velocity dispersion of F-type dwarfs in a
    simple harmonic oscillator potential with $\omega =
    69.1\pm1.1\kms\kpc\inv$, but this predicts the existence of an
    extended tail in the local velocity distribution (bottom middle
    panel).}\label{fig:gaia}
\end{figure*}

We define F-type dwarfs as those with near-infrared $J-K_s$ in the
range $0.143 < J-K_s < 0.3$. \citet{Bovy17a} have measured the
vertical stellar density profiles for different sub-types of F dwarfs
(\eg, F0V) from the \emph{TGAS} data, correcting for the selection
biases inherent in the \emph{TGAS} data. We use similar measurements
of the vertical stellar density of all F-type dwarfs (F0V through
F9V), defined as the combination of all of the sub-types considered by
\citet{Bovy17a}. These density measurements cover the range $-400\pc
\leq \zobs \leq 400\pc$ in $25\pc$ wide bins and are shown in the top
left panel of \figurename~\ref{fig:gaia}; $\zobs$ is the vertical
height as measured from the Sun's position, similar to the toy example
above.

We also measure the vertical velocity dispersion as a function of
vertical height from the \emph{TGAS} data. For this we select 103,603
F-type dwarfs using the same color and magnitude cuts as in
\citet{Bovy17a} and requiring relative parallax uncertainties less
than 10\,\%. These data provide us with $(v_{\alpha},v_{\delta}) =
(\mu_\alpha\cos\delta/\varpi,\mu_\delta/\varpi)$, where $\varpi$ is
the parallax and $(\mu_\alpha\cos\delta,\mu_\delta)$ are the proper
motion components in right ascension and declination. We obtain the
uncertainty covariance for each data point by Monte Carlo sampling
10,001 points from the correlated uncertainty covariance for the
parallax and proper motions. We fit the $v_z$ distribution from these
data by deconvolving the observed two-dimensional distribution of
$(v_{\alpha},v_{\delta})$ using a mixture-of-Gaussians model of the
velocity distribution in rectangular Galactic coordinates
$(v_x,v_y,v_z) = (U,V,W)$ using the extreme-deconvolution (XD;
\citealt{Bovy11a}) algorithm (see \citealt{Bovy09a} for a similar
application to \emph{Hipparcos} data). Because we are only interested
in the $v_z$ distribution and are not interested in the details of
this distribution, we use only two Gaussians. We fit this model in
$25\pc$ bins covering $-200\pc < \zobs < 200\pc$ and extract
$\sigma^2_z$. Outside of this $\zobs$ range, the data are too few and
the proper motions constrain $v_z$ too little to provide a useful
measurement of $\sigma^2_z$. We obtain uncertainties on these
$\sigma^2_z$ using 200 bootstrap resamplings. In the context of our
modeling we use these $\sigma^2_z$ measurements as a stand-in for
$\langle v_z^2\rangle$, that is, we assume that these have been
corrected for the solar motion. In principle we could marginalize over
the solar motion in the same way as we marginalize over the solar
position, but for the purpose of this illustration we will assume that
the correction for the solar motion is perfect. These data are shown
in the bottom left panel of \figurename~\ref{fig:gaia}.

We thus model the density $\nu_{\mathrm{obs}}(\tilde{z})$ and
mean-squared velocity $\langle v_z^2\rangle$. The latter is different
from the observable $\nu_{\mathrm{obs}}\langle
v_z^2\rangle(\tilde{z})$ that we have considered so far and requires
us to write down the various forces of change for the particle
weights, $\zsun$, and $\omega$ for the $\langle v_z^2\rangle$
observable. These forces of change are similar to the earlier
expressions, although they are slightly more complicated because the
weights enter into the normalization in the denominator. We give the
relevant expressions in the Appendix~\ref{sec:v2}. Because the
particle weights enter into the denominator of each $\langle
v_z^2\rangle$ measurement, the model is no longer linear in the
particle weights and the procedure for sampling the uncertainty
distribution of the particle weights is no longer strictly
correct. However, for large numbers of particle weights, the
normalization factor is only slightly affected by each individual
particle and the model is still close to linear in the particle
weights. We have run all of the mock tests described in the previous
sections for a mock data set consisting of density and $\langle
v_z^2\rangle$ measurements and find that the method proposed here
still works well. We thus apply it as is to the \emph{TGAS} data.

We use 10,000 $N$-body particles and start from a HO potential with a
frequency of $100\kms\kpc\inv$, an isothermal DF with $\sigma =
12\kms$, and a solar offset of $\zsun = 25\pc$. We use a kernel width
of $35\pc$. We then optimize the values of the particle weights,
\zsun, and the frequency $\omega$ using the observed \emph{TGAS} data
using 30,000 steps with $\Delta t \approx 0.1\Myr$ or a total time
$\approx 3\Gyr$. We use $\eps = 10^{-5.5}, \eps_{\odot} = 10^{-5}$,
and $\eps_\omega = 100$ and only update $\omega$ every 10 steps using
$\Delta \omega = 30\kms\kpc\inv$. We use a flat prior, $\mu = 0$.

The result from the M2M optimization is displayed in
\figurename~\ref{fig:gaia}. The M2M optimization quickly converges to
a well-constrained DF with $\zsun = -0.3\pc$ and $\omega =
69.8\kms\kpc\inv$. We run the MCMC algorithm for sampling the particle
weights, \zsun, and $\omega$ using 30,000 M2M optimization time steps
for sampling the weights and using a proposal distribution for $\zsun$
with width $\sigma_{\zsun} = 7\pc$ and a proposal for $\omega$ with
width $\sigma_\omega = 3\kms\kpc\inv$. We use 500 steps to average the
objective function for the MH steps for \zsun\ and 1,000 steps for
$\omega$, again changing $\omega$ to proposed values using 10,000
steps. We obtain 20 MH samples for $\zsun$ and 10 MH samples for
$\omega$ for each sample from the uncertainty distribution for the
particle weights. The MH acceptance fraction for $\zsun$ and $\omega$
was 0.25 and 0.27, respectively.

The resulting uncertainty in the observed density and $\langle
v_z^2\rangle$ as well as that in the inferred DF is shown in
\figurename~\ref{fig:gaia}. Because the density is so well measured,
the uncertainty in the model density is barely visible, but the
uncertainty in the kinematics is larger. The DF becomes uncertain at
large $z_{\mathrm{max}}$, but is well determined for orbits that stay
closer to the plane. Marginalizing over the uncertainty in the DF, we
find that $\zsun = -1\pm3\pc$ and $\omega = 69.1\pm1.1\kms\kpc\inv$.

The HO potential fits the data that we chose to model well. This is
surprising, because the local vertical potential should be quite
different from a constant density model (the HO model) over the
$800\pc$ range over which we have observed the density. The HO model
is able to fit the density data by having a large, high-energy
component in the DF, that is, the peak at $z_{\mathrm{max}} \approx
0.3\kpc$ in the top middle panel in \figurename~\ref{fig:gaia}. This
leads to two observable consequences in other panels of this figure:
the velocity dispersion increases for $|\zobs| \gtrsim 150\pc$ (bottom
left panel) and the local velocity distribution should display a wide,
high-velocity tail. An inspection of the \emph{TGAS} F-star kinematics
close to the plane where the vertical velocity is approximately given
by the vertical component of the proper motion shows that such a
high-velocity tail is absent in the observations (see also
\citealt{Holmberg00a}). This means that the HO potential is \emph{not}
a good model for the local vertical potential. Therefore, we do not
compare our constraint on $\omega$ to previous determinations of the
local gravitational potential \citep[\eg,][]{Holmberg00a} or interpret
our measurement of \zsun, which may be affected by the model for the
potential. Still, it is promising that the novel M2M algorithm
proposed in this paper works reasonably well with the observational
data with realistic uncertainties. We defer a more realistic treatment
of the vertical potential to future work.

\section{Discussion}\label{sec:discussion}

In the previous sections we have introduced various extensions of the
basic M2M method that are crucial to applying this method to model
observational data. Here we discuss the formal assumptions and
underpinnings of the sampling methods in more detail, comment on some
aspects of the method further, and describe other extensions and
improvements that could be made.

\subsection{On interpreting the M2M objective function as a PDF}\label{sec:m2maspdf}

The algorithm for sampling the uncertainty distribution of the
particle weights and the MCMC algorithms for sampling nuisance and
potential parameters depend on our assumption that we can interpret
the M2M objective function as the logarithm of a PDF for the
parameters. However, the M2M objective function, defined in
\equationname~\eqref{eq:objective}, is not a static function, but
fluctuates as the M2M particles orbit, even when all parameters are
held fixed. Thus, the interpretation of the M2M objective function as
a PDF is not obvious. We argue now that when run properly, the M2M
procedure optimizes and samples from a well-defined, correct PDF.

M2M modeling can be seen as an approximation of Schwarzschild
modeling. Schwarzschild modeling uses the same form of the objective
function, except that the kernels that in M2M are evaluated for a
snapshot are in Schwarzschild modeling averaged in time. The objective
function in that case defines the logarithm of a well-defined, static
PDF. \citet{Malvido15a} have shown that M2M optimization is formally
equivalent to Schwarzschild optimization in the limit of large times,
large $N$, and small $\eps$. Therefore, the basic M2M optimization
procedure in fact optimizes a well-defined, static objective function
if the optimization is performed sufficiently slowly, that is, over a
long enough time and with small $\eps$. Moreover, our proposed
sampling procedure for the uncertainty in the particle weights also
optimizes the same objective function and thus effectively samples the
well-defined, Schwarzschild PDF. To the extent that the objective
function is convex (exactly so for the objective function in our mock
example above when no smoothing is applied), there is also no danger
of optimizing to a local maximum.

To sample parameters other than the particle weights we have
introduced Metropolis-Hastings algorithms that use the averaged
objective function as the logarithm of the PDF. The correct objective
function is once again the equivalent Schwarzschild objective
function. The question is in what limit these two are equivalent. For
a single observation $Y$, we can schematically write down the
contribution to $\chi^2$ as (ignoring the observational uncertainty in
the denominator)
\begin{equation}\label{eq:m2m-ex}
  \chi^2_{\mathrm{M2M}} = \left(\sum_i \wi K_i - Y\right)^2\,,
\end{equation}
where $K_i$ are the relevant kernel functions. The equivalent
Schwarzschild form of this equation is 
\begin{equation}
  \chi^2_{\mathrm{Schwarzschild}} = \left(\sum_i \wi \langle K_i\rangle - Y\right)^2\,,
\end{equation}
where $\langle K_i\rangle$ denotes the orbit-averaged
kernel. Averaging \equationname~\eqref{eq:m2m-ex} gives
\begin{align}\label{eq:m2m-as-schwarzschild}
  \begin{split}
  \langle \chi^2_{\mathrm{M2M}} \rangle & = \sum_{i,j} \wi\,w_j\langle K_i K_j\rangle - 2 Y \sum_i \wi \langle K_i \rangle +  Y^2\,,\\
  & = \sum_{i,j} \wi\,w_j \rho_{ij}\,\sigma_{K_i}\,\sigma_{K_j} + \chi^2_{\mathrm{Schwarzschild}}\,,
\end{split}
\end{align}
where $\rho_{ij}$ is the correlation matrix of the orbital kernels:
$\rho_{ij}\,\sigma_{K_i}\,\sigma_{K_j} = \langle (K_i-\langle
K_i\rangle)\,(K_j-\langle K_j\rangle)\rangle$ and $\sigma_{K_i} =
\sqrt{\langle (K_i-\langle K_i\rangle)^2\rangle}$. Thus, for the
orbit-averaged M2M objective function to be a good approximation to
the Schwarzschild objective function, we need
\begin{equation}
\sum_{i,j} \wi\,w_j \rho_{ij}\,\sigma_{K_i}\,\sigma_{K_j} \ll \chi^2_{\mathrm{Schwarzschild}}\,.
\end{equation}
Orbits with very different trajectories have $\rho_{ij} \approx 0$,
while orbits with similar trajectories have $\wi \approx w_j$ and $K_i
\approx K_j$. Therefore, we can simplify the left-hand side of the
previous equation to a sum over sets of orbits with similar
trajectories
\begin{equation}
\sum_{i,j} \wi\,w_j \rho_{ij}\,\sigma_{K_i}\,\sigma_{K_j} \approx
\sum_{\mathrm{sets\ of\ orbits}\ i}\,\wi^2\,\sigma^2_{K_i}\,\sum_{j\ \mathrm{similar\ to}\ i}
\rho_{ij}\,.
\end{equation}
For a large enough number of M2M particles distributed randomly in
orbital phase, $\sum_{j\ \mathrm{similar\ to}\ i} \rho_{ij} \approx
0$. Thus, if the M2M system consists of a large number $N$ of
particles with well-mixed phases, the averaged M2M objective function
is a good approximation to the Schwarzschild objective function and
can therefore be used as the logarithm of the PDF in a
Metropolis-Hastings update.

\subsection{On the approximate data-resampling technique for sampling the particle-weights PDF}\label{sec:discuss_weights}

As we already stressed in \sectionname~\ref{sec:wunc}, the method for
sampling the particle-weights PDF by resampling the data and obtaining
the best-fit weights for the resampled data is approximate in the
sense that it technically only applies in the case that the particle
weights can take any value, both positive and negative. When the
weights are required to be positive, the data-resampling technique
will oversample the zero-weight edge of parameter space---oversample
it relative to its correct proportion under the PDF. Because a
non-parametric method such as M2M can only constrain the gravitational
potential by excluding solutions that require particles to have
negative weights---without this restriction, the data can always be
perfectly fit---this is a matter of concern. Here we provide some
arguments that demonstrate that this is not a major problem.

There are two general statements that we can make about any M2M
particle-weights PDF. The first is that for gravitational potentials
that are close to the true potential, all orbits should be able to
contribute non-negatively to the DF. Some orbital families may not
exist and thus be constrained to have weights close to zero, but no
orbits should require negative weights. This means in particular that
we may assume that the mode of the PDF lies in the volume of
$\mathbb{R}^N$ where all weights are non-negative (if the weights are
underconstrained, the mode is a trough that for the correct potential
will extend into this volume; we will continue to use `mode' below,
but this always includes the `trough' case as well). A second general
property of the M2M PDF for the weights is that it is log-convex if
the uncertainties are Gaussian and the predicted observables depend
linearly on the particle weights. Combined with the fact that the mode
of the PDF has all weights non-negative for gravitational potentials
close to the true one, this implies that a significant amount of
probability mass has non-negative weights.

We use the data-resampling technique for sampling the particle-weights
PDF for two purposes: (i) for sampling the uncertainty in the particle
DF when investigating the DF and (ii) to marginalize over the
uncertainty in the particle DF when constraining the gravitational
potential or any nuisance parameters.  When using the data-resampling
technique for studying the uncertainty in the orbital DF, presumably
this will be done for a reasonable gravitational potential (otherwise
one would adjust the potential first). The data-resampling technique
will oversample weights exactly equal to zero for those orbits that
are required to have negative weights to provide a good fit to the
observations. The arguments in the previous paragraph demonstrate that
this oversampling will only have a minor effect, because the mode is
expected to have all non-negative particle weights.

In the full MCMC method for constraining nuisance parameters and the
parameters of an external gravitational potential, the data-resampling
technique is used to marginalize over the uncertainty in the particle
DF. The MH steps are based on the ratio of the PDF for the current and
the proposed set of parameters. That the data-resampling method
oversamples weights being exactly equal to zero has a different effect
based on whether the mode of the weights PDF has all non-negative
weights (the case for well-fitting potentials) or whether it wants to
have some negative weights (the case for badly-fitting potentials). In
the first case, the likelihood of the model will be biased low by the
oversampling, because there is a higher chance than there should be
(under the correct PDF) of sampling the zero-weight edge, where the
PDF has a lower value than in the $\wi > 0$ region (this follows from
the log-convexity of the likelihood). In the second case, the
likelihood of the model will be biased high, because the oversampled
edge now has higher probability than the $\wi > 0$ region. Thus, the
MH sampling is biased toward worse-fitting models and the
data-resampling technique will therefore lead to conservative,
somewhat inflated uncertainties on the parameters of the
model. However, the more severe the oversampling of zero weights, the
worse-fitting the model must be in the first place, because the
oversampling fraction---the fraction of samples that land on the
zero-weight edge---gets larger the further the mode of the
particle-weights PDF is from the non-negative part of parameter space
(the oversampling fraction is equal to the integral of the PDF over
the part of space where at least one $\wi$ is negative when the $\wi$
are allowed to take on negative values, divided by the integral of the
PDF over all space; for this to be large for a Gaussian PDF the mode
needs to be deep in the negative-weight region). Therefore, the bias
will be small, because even when evaluated too often with weights set
to zero, these models will still have low likelihoods relative to
better fitting models. We see no evidence of significantly inflated
uncertainties on the nuisance or gravitational-potential parameters
from the limitations of the data-resampling technique in any of the
experiments in this paper.

\subsection{Aspects of the method}\label{sec:discuss_aspects}

\emph{Fixing the sum of the particle weights:} From when M2M was first
proposed, the sum of the particle weights has typically been fixed to
a constant, under the assumption that the total mass of the modeled
system is known. The standard M2M algorithm does not conserve the sum
of the particle weights and the weights are typically simply
renormalized after each update step. We have left the sum of the
particle weights free to be constrained by the data, which is the
appropriate thing to do because the total mass is never exactly
known. This completely gets around the issue of the weights
renormalization. Nevertheless, when setting up an $N$-body simulation
using the M2M method one may want to constrain the total mass of the
system to a specific value. A simple way to do this is to (a) define
the particle weights to sum to one, in which case the weights cover
the simplex embedded in $N$-dimensional space, and (b) transform the
simplex to a $N-1$ dimensional space that covers all of
$\mathbb{R}^{N-1}$. We discuss how to do this in
\appendixname~\ref{sec:simplex}.

\emph{The importance of the integration method:} We have sidestepped
the issue of orbit integration in our example of a HO potential,
because orbit integration can be performed analytically in this
model. However, in more realistic models, orbits need to be integrated
numerically with a small enough time step such that numerical errors
are small. While typically not important in galactic dynamics, we
recommend use of a symplectic integrator such as leapfrog for the
following reason. When performing the entire sampling procedure,
orbits can be integrated for thousands of dynamical times or more and
small energy errors can accumulate to a significant fraction of the
energy. Symplectic integrators with a small time step typically avoid
growth of the energy error and faithfully follow the evolution of the
dynamical system being modeled over many dynamical times.

\emph{Other MCMC samplers:} In algorithms~\ref{alg:nuisance} and
~\ref{alg:potential} we have opted to use a simple Metropolis-Hastings
sampler to sample the nuisance and potential parameters. However, in
applications with more nuisance parameters or more complicated
potential models, we may want to use a MCMC sampler that is less
sensitive to the proposal step size or explores the PDF more
efficiently. Of particular interest is Hamiltonian Monte Carlo
\citep{Duane87a,Neal11a}, which can make large strides across the PDF
by making use of the derivatives of the PDF. For the nuisance
parameters, we can straightforwardly compute these derivative as the
average force of change similar to how the average objective function
is computed in algorithm~\ref{alg:nuisance}.

\subsection{Directions for future work}\label{sec:discuss_future}

\emph{Self-consistently generating the potential:} One attractive
aspect of M2M modeling compared to other dynamical modeling approaches
is that it is possible to let the M2M particles generate the
gravitational force field or some part of it
\citep[\eg,][]{Hunt13a}. That is, when modeling the stellar kinematics
of, for example, an external galaxy, one can run the M2M optimization
while integrating the particles in the gravitational potential that
they themselves generate (plus perhaps additional dark matter). If the
particle weights are changed slowly enough, the potential changes
adiabatically and if the number of particles is large enough, the
potential, being the combination of many particles, changes on longer
timescales than the individual particle weights. Therefore, the
arguments above that demonstrate that the M2M procedure optimizes a
well-defined objective function still hold. The data-resampling method
for obtaining uncertainties on the values of the particle weights
should therefore still perform well. In the MCMC updates of the
nuisance parameters, the particle weights are held fixed and the
gravitational force generated by the particles should therefore not
change much (it could be held fixed). In the MCMC updates for the
parameters of the external gravitational potential, the orbits are
changed adiabatically and the potential generated by the particles
needs to be updated on the fly as well to preserve the consistency
between the M2M particles and the potential.

\emph{Dynamical stability:} When we do not demand that the M2M
particles generate (part of) the gravitational potential, one can end
up with a solution or an MCMC sample that is dynamically unstable. M2M
modeling, by virtue of using particles, can easily add the constraint
of dynamical stability after the fact, by using the set of particle
weights for a given MCMC sample to initialize an $N$-body simulation
and determining whether it is dynamically stable or not. Samples that
are not stable could be rejected and pruned from the chain.

\emph{Priors on the particle weights:} We have paid little attention
to the penalization term (the prior) in the M2M objective function and
set it to zero in all of our examples (corresponding to an improper,
flat prior on the weights). While it is clear that we do have definite
prior beliefs about the particle weights, these are not well expressed
by the standard entropy-like M2M or Schwarzschild penalization terms
in the objective function. These standard forms express the prior
belief that the particle weights are close to a reference set of
weights, but without any correlation between the weights of similar
orbits. This is problematic when we want to sample the uncertainty
distribution of the particle weights. Interpreting the standard
penalization as the logarithm of a prior PDF and sampling from this
prior PDF gives sets of particle weights in which similar orbits can
have widely different weights. A better prior would express the fact
that similar orbits have similar weights without necessarily having
strong prior beliefs about the actual value of the weights. This
could, for example, be done using a Gaussian process with a kernel
function in the space of integrals of the motion. Alternatively, a
local smoothing of the current set of particle weights could be
substituted for the prior \citep{Morganti12a}. One advantage of using
a Gaussian process is that this would allow the prior to be taken into
account in the data-resampling technique for sampling the uncertainty
in the particle weights: we can `resample' the mean of the prior
applied in each optimization sequence similar to how each data point
is resampled in this technique and this returns formally correct
samples from the posterior PDF for the particle weights (as long as
they are positive). For spherical or axisymmetric systems integrals of
the motion are available that can be used to evaluate the similarity
of orbits, but even in general time-independent systems the energy
could be used or one can construct other similarity functions.

\emph{Modeling multiple populations:} In our mock example, we have
assumed that only a single population of stars is being
modeled. However, if density and kinematics measurements are available
for different populations of stars, one could use the same set of
particles with multiple weights associated with each particle, one for
each stellar population. That is, suppose that we had modeled both F
and G-type dwarfs in \emph{Gaia} DR1 as an example, we could have used
$N$ particles with two weights for each particle, one for F-type stars
and one for G-type stars. These weights can all be optimized
simultaneously. More generally, if we have additional information such
as overall metallicity $Z$, abundance ratios, or ages for stars, we
can replace the particle weights $\wi$ associated with each particle
with parameterized functions, \eg, $\wi(Z)$, of these additional
quantities and fit for the parameters of these functions. One
particularly attractive way of doing this is to represent these
functions in terms of basis functions with free amplitude parameters,
\eg, $\wi(Z) = \sum_k \alpha_{ik}\,\beta_k(Z)$ with $\beta_k(\cdot)$ a
set of fixed basis functions. In this case, the observables remain
linearly related to the parameters ($\alpha_{ik}$) and the
data-resampling technique for obtaining uncertainties on the particle
weights then also applies to the amplitudes of the basis functions.

\section{Conclusion}\label{sec:conclusion}

M2M modeling is one of the most promising dynamical-modeling methods
for fitting observational constraints on relaxed stellar systems
without making additional assumptions about the shape of the system's
DF. This generality is a prerequisite to making the most robust
inferences regarding the stellar, baryonic, and dark masses of stellar
systems. M2M has been used successfully to model the dynamics of
external galaxies \citep[\eg,][]{DeLorenzi08a} and of the bar-shaped
inner Milky-Way region \citep[\eg,][]{Portail17a}. However, so far M2M
models (or Schwarzschild models for that matter;
\citealt{Magorrian06a}) have not dealt with the massive degeneracies
that necessarily accompany a DF model as flexible as that used in
M2M. Because these degeneracies can have a large influence on the
inferences about the gravitational potential made using M2M modeling,
results obtained without taking the uncertainty in the particle
distribution into account should be viewed with suspicion.

We have improved and extended the standard M2M algorithm for fitting
observational data in various ways. Firstly, we have shown that all
parameters describing the system---particle weights, nuisance
parameters, and the parameters of an external gravitational
field---can be optimized simultaneously in the M2M optimization. This
makes it much easier to fit M2M models to observational data, as only
a single M2M run is necessary, no matter how complicated the nuisance
parameters or external gravitational potential is.

Secondly, we have introduced algorithms to sample from the full
posterior PDF that describes the uncertainty in the particle weights
and the nuisance and gravitational-potential parameters. For the
particle weights, which can be very numerous, this is done through a
technique that resamples the data within its uncertainties and turns
the sampling problem into an optimization problem. This technique is
formally correct when the model is linear in the parameters and the
data uncertainties are Gaussian. This is typically the case for M2M,
where the model typically consists of kernels combined using linear
weights, but we have also shown that this techniques works when the
data is the second moment of the velocity distribution. We sample the
nuisance parameters and those describing the external gravitational
field through a carefully designed Metropolis-Hastings MCMC algorithm,
where the averaged M2M objective function is used as the logarithm of
the PDF and the potential is only ever changed adiabatically. Because
of the tight connection between M2M and Schwarzschild modeling
demonstrated in \citet{Malvido15a} and in
\sectionname~\ref{sec:m2maspdf} these new techniques can also be
useful for Schwarzschild modeling.

The full M2M method described in this paper allows for large-scale,
fully-probabilistic modeling of observational data. It will be useful
in future modeling of data on Milky-Way stars \citep[\eg,][]{Hunt14a}
and on external galaxies. As a first example, we have analyzed data on
the vertical density and kinematics of F-type dwarfs from \emph{Gaia}
DR1 in a simple harmonic-oscillator model for the local gravitational
potential. We find that we can fit the data that we have chosen to
model, but a more realistic model for the gravitational potential is
necessary to make definitive statements about what these data imply
about the local mass distribution.

All of the analysis in this paper
can be reproduced using the code found
at\\ \centerline{\url{https://github.com/jobovy/simple-m2m}}.

{\bf Acknowledgements} We thank the anonymous referee for a
constructive report. JB received support from the Natural Sciences and
Engineering Research Council of Canada. JB also received partial
support from an Alfred P. Sloan Fellowship and from the Simons
Foundation. DK acknowledges the support of the UK's Science \&
Technology Facilities Council (STFC Grant ST/K000977/1 and
ST/N000811/1). This work has made use of data from the European Space
Agency (ESA) mission {\it Gaia}
(\mbox{\url{http://www.cosmos.esa.int/gaia}}), processed by the {\it
  Gaia} Data Processing and Analysis Consortium (DPAC,
\mbox{\url{http://www.cosmos.esa.int/web/gaia/dpac/consortium}}).

\appendix

\appendix

\section{Using the mean-squared velocity as the observable}\label{sec:v2}

Instead of the density-weighted mean-squared velocity shown in
equation~\eqref{eq:modeldensv2}, we can use the mean-squared velocity,
$\langle \vz^2\rangle(\zobs_j)$, itself as an observable. This allows
us to use the velocity measurements from a sub-sample of the one used
for the density measurement. The model mean-squared velocity is
defined as
\begin{equation}\label{eq:modeldv2}
  \langle \vz^2\rangle(\zobs_j) = \sum_i \wi \vzi^2\,K^0(|\zobs_j+\zsun - \zi|;h)/\dens_{v^2,j}\,, 
  \end{equation}
where $\dens_{v^2,j}=\sum_i \wi K_j^0(\zi;h)$, corresponding to a
choice of a kernel of $K_j^{v^2}(\zi,\vzi) =
\vzi^2\,K_j^0(\zi;h)/\dens_{v^2,j}$. Note that the denominator can
be calculated using a different kernel (or kernel width) than the
density itself (\equationname~[\ref{eq:modeldens}]) and, therefore, we
use $\dens_{v^2,j}$ which can be different from
$\dens(\zobs_j)$. Assuming that the $\langle \vz^2\rangle(\zobs_j)$
observations have a Gaussian error distribution with variance
$\sigma^2_{v,j}$, the contribution to $\chi^2$ from $\langle
\vz^2\rangle(\zobs_j)$ is given by
\begin{equation}
  \chi_{j,v^2}^2 = [\Delta^{v^2}_j/\sigma_{2,j}]^2 = \left( \langle \vz^2\rangle(\zobs_j)-\langle \vz^2\rangle^\mathrm{obs}_j\right)^2/\sigma^2_{v,j}\,.
\end{equation}
In this case, the contribution from $\langle \vz^2\rangle(\zobs_j)$ to
the force of change for the particle weights becomes
\begin{align}\label{eq:fcwv2}
-\frac{1}{2}\frac{\partial \chi^2_{j,v^2}}{\partial \wi} = & -\frac{\Delta^{v^2}_j[\vzi^2-\langle \vz^2\rangle(\zobs_j)]\,K^0_j(\zi;h)}{\sigma^2_{v,j}\,\dens_{v^2,j}}\,.
\end{align}

Similarly, the contribution from $\langle \vz^2\rangle(\zobs_j)$ to
the force of change for \zsun\ is
\begin{align}\label{eq.vz2it}
- &\frac{1}{2} \, \frac{\partial \chi^2_{j,v^2}}{\partial \zsun} = 
 - \frac{\Delta^{v^2}_j}{\sigma^2_{v,j}\,\dens_{v^2,j}}\\& \ \times \mathlarger{\mathlarger{\sum}}_{i} \wi\,[\vzi^2-\langle \vz^2\rangle(\zobs_j)]\, \frac{\dd K^0_j(r;h)}{\dd r}\Bigg|_{|\zobs_j+\zsun-\zi|}\!\!\!\!\!\!\!\!\!\!\!\!\!\!\!\!\!\!\!\!\!\!\!\!\mathrm{sign}(\zobs_j+\zsun-\zi)\nonumber\,.
\end{align}

The force-of-change for $\omega$ is again computed using a direct
finite difference, similar to \equationname~\eqref{eq:fcomega}.

\section{M2M on the simplex}\label{sec:simplex}

If one wants to run M2M modeling under a hard constraint on the sum of
the particle weights (\eg, if the total mass represented by the M2M
particles is exactly known, as in setting up an $N$-body simulation),
the standard M2M force-of-change-based algorithm fails because the
update equations for the particle weights do not conserve the sum of
the weights. No satisfactory solution of this problem has been
proposed in the literature.

If the particle weights must sum to a constant value we can always
redefine them such that they sum to one: $\sum_i \wi = 1$. The weights
are then constrained to be positive and to sum to one and they
therefore define a $N-1$ dimensional simplex embedded in
$\mathbb{R}^N$. We can then re-write the M2M algorithm in terms of a
transformed set of variables $\yi$ that cover all of
$\mathbb{R}^{N-1}$ and that parameterize the simplex. In this case,
the particle weights always exactly sum to one and the algorithm
cannot stray from this condition. Generically, such a transformation
would require $\mathcal{O}(N^2)$ operations to compute the derivatives
with respect to the $\yi$ from those with respect to the $\wi$. Here
we propose a specific transformation that is simple to implement and
for which the derivatives with respect to $\yi$ can be computed in
$\mathcal{O}(N)$ time. Transforming to the $\yi$ is then a feasible
method even for very large numbers of $N$-body particles.

The transformation from $\wi$ to $\yi$ is the combination of the
following transformations (partially following
\citealt{Betancourt12a})
\begin{align}
  x_i & = 1-\frac{\wi}{\prod_{k=1}^{i-1} x_k}\,,\\
  \yi & = \mathrm{logit}(x_i)-\mathrm{logit}(X_N)\,,
\end{align}
where $\mathrm{logit}(\cdot)$ is the log-odds function
$\mathrm{logit}(x) = \ln\left(x/[1-x]\right)$ with the inverse
$\mathrm{logit}^{-1}(x) = 1/[1+e^{-x}]$. $X_N$ is a $N-1$ dimensional
vector with entries
$[\frac{N-1}{N},\frac{N-2}{N-1},\ldots,\frac{1}{2}]$, which causes the
simplex with all particle weights equal to each other, $\wi = 1/N$, to
be mapped to the zero vector. The inverse transformation is given by
\begin{align}\label{eq:invtransform}
  x_i & = \mathrm{logit}^{-1}(\yi+\mathrm{logit}(X_N))\,,\\
  \wi & = \left(\prod_{k=1}^{i-1} x_k\right)\cdot\begin{cases}
  1-x_i, & i < N \\
      1, & i = N
    \end{cases}\,.\label{eq:invtransform2}
\end{align}
This inverse transformation is straightforward to implement using
vectorized operations, while the $\wi \rightarrow \yi$ transformation
requires a loop to accumulate the product in the first line. The
inverse transformation is the one that is relevant for evaluating the
objective function during the running of the M2M algorithm. The $\wi
\rightarrow \yi$ transformation is only needed at initialization (if
the weights are initialized as $\wi = 1/N$, then the initial $y_i = 0$
for all $i$).

To run the M2M algorithm in terms of the $\yi$ variables, we 
compute the derivative of the objective function $F$ using the chain rule. The Jacobian $\partial w_k / \partial y_i$ of this transformation is
\citep[cf.][]{Betancourt12a}
\begin{equation}
  \frac{\partial w_k}{\partial y_i} = \begin{cases}
  w_k\,(1-x_i), & i < k \\
  -\wi\,x_i, & i = k\\
  0, & i > k
    \end{cases}\,.
\end{equation}
This is a lower-triangular matrix. The chain rule can then be
simplified to
\begin{align}
\begin{split}
  \frac{\partial F}{\partial \yi} 
  & = -x_i \wi \frac{\partial F}{\partial w_i} + (1-x_i)\,\sum_{k=i+1}^{N} w_k\,\frac{\partial F}{\partial w_k}\,.
\end{split}
\end{align}
All $N-1$ derivatives can be computed together in $\mathcal{O}(N)$
time by accumulating the sum.

If one interprets the objective function as the logarithm of a
probability distribution, transforming to a new set of variables
requires tracking the determinant of the Jacobian. Because the
Jacobian is a lower-triangular matrix, its determinant is given by the
product of the diagonal entries
\begin{equation}\label{eq:logdet}
  \left|\frac{\partial w}{\partial y}\right| = \prod_{k=1}^{N-1}
  w_k\,x_k\,.
\end{equation}
The derivative of the logarithm of the Jacobian with respect to $\yi$
is given by
\begin{equation}
  \frac{\partial}{\partial \yi}\ln\left|\frac{\partial w}{\partial y}\right|
  = (N-i)\,(1-x_i)-x_i\,,
\end{equation}
for $i=1,\ldots,N-1$. 

\end{document}